UNIVERSITY OF CALIFORNIA, MERCED

# Multi Query Optimization in GLADE

A thesis submitted in partial satisfaction of the

requirements for the degree

Master of Science

in

Electrical Engineering and Computer Science

by

Abdur Rafay

Committee in charge:

Prof. Florin Rusu, Chair
Prof. Mukesh Singhal
Prof. Sungjin Im

2016

The thesis of Abdur Rafay is approved, and it is acceptable in quality and form for publication on microfilm and electronically:

Chair

University of California, Merced

2016

# TABLE OF CONTENTS

# DECLARATION

I hereby declare that no portion of the work referred to in this Project Thesis has been submitted in support of an application for another degree or qualification of this of any other university or other institute of learning. If any act of plagiarism found, I am fully responsible for every disciplinary action taken against me depending upon the seriousness of the proven offence.

# ACKNOWLEDGEMENTS

First and foremost I would like to thank the Almighty, who showed this project the light of the day and enabled me to achieve the targets and goals I had set in my mind before going for the project.

I would like to thank my parents whose prayers and support was always there to raise my morale in difficult times and who have always been a source of inspiration and guidance throughout my educational career.

I would also like to extend my heartfelt thanks to my project advisor Professor Florin Rusu, who was always there for me whenever I needed him and has been a source of knowledge and guidance in the hours of need, without the guidance of whom this project wouldn't have been possible.

# ABSTRACT

**Multi Query Optimization in GLADE**

by

Abdur Rafay

Master of Science

in

Electrical Engineering and Computer Science

University of California, Merced
2016

SQL-on-Hadoop systems, query optimization, data distribution over multiple nodes and parallelization techniques are few of the areas under extreme research these days. Big names like Amazon, Google, Microsoft and many more are working on implementing systems for faster access of data from multiple nodes, reducing data mobility and increasing the parallelization. Customer's queries are retrieved and reviewed by the database systems in an efficient way in the least amount of time by the introduction of various parallelization techniques by running the same query in parallel over different nodes carrying the data. Apart from multi-threading parallelization, there is another way of parallelization that can be performed in order to further reduce retrieval time, hence improving the efficiency of the system; parallelization on user queries on top of a DBMS/RDBMS. In this paper, we will study one such technique of how multiple queries can run simultaneously on a system in order to increase the system efficiency by reducing the data retrieval from the storage. Maximum sharing of workload has been performed by generating optimal and ubiquitous join plans for a set of queries and then fed them to GLADE (Generalized Linear Aggregate Distributed Engine), a scalable distributed system for large scale data analytics. Our main work is centered on generating GLADE join plans for a Multi-Query, satisfying maximum number of queries in order to maximize data sharing and minimize data retrieval for each individual query.

# Chapter 1

# Introduction

## 1.1 SQL and DBMS

Programming lanuages or machine languages helps us interact with machines. There are different types of programming languages in order to communicate with the machine and pass instructions to get our work done in an efficient way. We have static and dynamic programming languages, which are then further breakdown into weak and strong languages (the detail of this is beyond the scope of this paper). While we are talking to machines in different languages instructing them to perform our tasks, we need some type of storage buffer, where we can store data in order to fetch it later on, when needed. This DATABASE system is a place where the data is stored, in a structured or unstructured format, organized or randomly spread, depending upon the type and need of data. The data, if stored in a structured manner in the database, is stored in the form of tables. When there is a need of storing a data in the form of database, there came the need of implementing a system, which can provide the user or the customers to interact with the database engine, where data is stored, and perform different tasks on the user stored data. Such systems are called Database Management Systems (DBMS).

DBMS provide the user to interact with the computer via some environment which can be a GUI (graphical user interface) or the form of passing the commands and talking to the computer using some special syntax. This special syntax is actually a designated language assigned to communicate with the database, and is called Structured Query Language or SQL. Via SQL, we have full control of our database and can fetch or add more data, update the existing data and so on.

There are different types of existing DBMS's like Oracle, IBM DB2, Microsoft SQL server, Access, MySQL, Sybase and many more. On the other hand, SQL is further categorized into 3 different types of command sets, namely data definition language, data manipulation language, and data control language. The most standard SQL commands with which one can almost communicate completely with the database system are CREATE, INSERT, SELECT, UPDATE, DROP and DELETE.

## 1.2 RDBMS

RDBMS stands for relational DBMS, which is a DBMS that contains data stored in form of tables and the tables are being related to each other through some attribute(s). The power of RDBMS is its property of relation among the tables, because of which we can break down bigger tables into smaller tables with some common relation between them. As the size of the data is growing exponentially, the need of RDBMS is increasing, since we don't end up having such a huge amount of memory at one place, hence we cannot store all the data at one place. So the ease provided by RDBMS gets handy here. We will talk about GLADE in this paper shortly and explain how the testing is being done in GLADE using RDBMS.

## 1.3 SQL-on-Hadoop Hype

In the previous section, we talked about structured query language. In this section, we will put some light on Hadoop and then SQL-on-Hadoop systems. Hadoop or Apache Hadoop is an open source software framework that lets you store your data in large amount (of the order of $10^{15}$ and more) over a distributed system with multiple clusters, and then later on let you process such big data using simple programming models. It comes in handy when the data is extremely large and you cannot store it on a single machine. Hadoop distributed file system (HDFS) is an extremely fault tolerant and low cost file system that provides scalable and reliable data storage and control on your stored data. The data stored is in terabytes to petabytes in size and even more.

So what is the hype about systems with SQL on top of Hadoop or HDFS? These days, companies are increasingly inclining towards using Apache Hadoop as the underlying system to store their data. On the other hand, to manage the data, they deeply rely on SQL because of the familiarity and ease of using SQL. As a result over the past few decades, by supporting familiar SQL queries, SQL processing over HDFS is getting more and more attention, and number of enterprise developers and business analysts providing such SQL-on-Hadoop functionality in their architecture is increasing exponentially, GLADE being one of them. GLADE used to work on a different M4 language format for the queries, but we will get to that later on in the paper on how SQL support is being introduced in GLADE.

Clearly, next generation of big data is coming in full bloom to help enterprises and Business Intelligence (BI) to store, process and analyze big data more effectively and efficiently.

## 1.4 GLADE Introduction

GLADE is an acronym for Generalized Linear Aggregate (GLA) Distributed Engine and is a scalable distributed framework for large scale data analytics. All the implementation of the algorithms in the paper are done in GLADE and all the queries were tested in GLADE too. We will discuss in depth about the architecture of GLADE and how it works in Chapter 4. GLADE, a relational execution engine, basically consists of a simple user interface to create user defined functions, here we call them GLA's, the heart of GLADE, and a distributed system that executes these GLA's in parallel. We will discuss more about how different type of parallelism is used in GLADE. It is derived from DataPath [2] which is an extremely efficient multi-query processing framework.

## 1.5 Problem Formulation

Query execution in a distributed system is one of the hot topic these days. BI are working on making their systems more efficient trying to execute queries as fast as possible and as parallel as possible to achieve lesser time and process more data. One mundane way of achieving faster data retrieval is parallelization among the nodes. In this paper, we will consider another rather less common but more efficient way of parallelization on queries, on top on node parallelization. We call it as Multi Query Optimization, which is discussed in detail in Chapter 5.

## 1.6 Solution Analysis

In this paper, we will show that how parallelization is achieved by running the same query on all the nodes having the data, and how same execution plan runs on each node to extract data. All the nodes passes the data back to one master node, where the data aggregation is done and the results are passed back to the user. This is one of the monotonous way of query execution. In this paper, we will consider another way of parallelization on queries. In that case, our system will create chunks of queries and call each chunk as a Multi Query (MQ).

A MQ consists of a set of queries, where each query contains certain distinct tables. There is only one constraint on how each MQ is defined, and that is there has to be one query in a MQ, that consists of all the tables used in all the other queries collectively. Each MQ is processed to generate a single shared GLADE join plan, which is a graph with multiple exit points. Each exit point correspond to each query in a MQ. The whole graph is executed on all the nodes to generate results for all the queries simultaneously.

The main idea behind this algorithm is to achieve faster retrieval of data from the underlying system without the need of accessing it multiple number of times. This redundancy is simply removed by pairing the same set of queries sharing the same tables. For example if we have a MQ with $q$ number of queries, and say $c(q')$ corresponds to the worst case cost of the query containing all the tables for this particular MQ, then the overall cost for all the queries collectively as a MQ is upper bounded by $\lceil c(q') \rceil$. The main algorithm will be discussed in the Chapter 5 and we will show how this algorithm is derived from [1].

## 1.7 Existing solutions and their limitations

There has been a lot of work going on in the area of MQ optimization. BI are trying to optimize their distributed systems by implementing various algorithms, trying to reduce the retrieval time and get the results as fast as possible. The fight for creating a perfect distributed system that generates millions and millions of tuples within microseconds is real. [1], [3] and [4] have done a lot of work on optimizing the cost of running multiple queries. In this paper, we will discuss about a new technique of parallelization, parallelization of queries which we named as Multi Query Optimization. The idea is somewhat similar to [1] but with modifications and considering sharing of queries in a different way, by incorporating satisfied and unsatisfied queries while generating GLADE join plans.

## 1.6 Contributions

GLADE has already been on the play from past 5 years and advancements are being carried on in it. The main contributions we brought into GLADE are providing support to handle multiple queries at a time. We also provide the support for SQL queries in GLADE, which was previously working on M4 format. We introduce the keyword MULTIQUERY in GLADE's dictionary of SQL (more detail on that in Chapter 5) and performed different tests for different set of MQ's in GLADE. To abridge, here is the list of all the contributions in this paper that we have introduced in GLADE:

- ✓ Integrate SQL query parser, query compiler and query optimizer in GLADE.
- ✓ Implement support for SQL queries into GLADE by converting SQL text to GLADE execution plans.
- ✓ Implement support for multi-queries in GLADE using Yacc.

## 1.7 Report Timeline

The report is divided into 6 chapters. We have already discussed the introduction chapter. In the next chapter, we will discuss about query planning and query execution in general, how GLADE plans are generated and what are different components of query planner. In chapter 3, we will discuss how we incorporated query optimization in GLADE to generate execution plans to run on GLADE nodes and how we provided SQL support in GLADE. In chapter 4, we will discuss about GLADE and its architecture in general. Chapter 5 finally talks about multi query optimization algorithm in GLADE on top of simple query optimizer, which is a variant of [1]. Chapter 6 concludes our paper.

# Chapter 2

# Query Planning

## 2.1 Introduction

Database management systems helps us to interact with the database where the data is stored somewhere in the underlying memory. This is a big picture, and in order to understand what is happening at a lower level, we need to get inside on how queries are being processed inside the engine and what is the most optimal way to retrieve the data from the storage. Query planner tells us how a relational DBMS evaluates and processes the queries, how different components of DBMS work together to take a query from a user and parse it, generate join plans consisting of joins of the tables from which the data is to be fetched, and eventually execute those join plans to generate tuples for the user.

There are 4 basic steps in which a query is being processed in GLADE:

- Query Parsing
- Query Optimization
- Query Compilation
- Query Execution

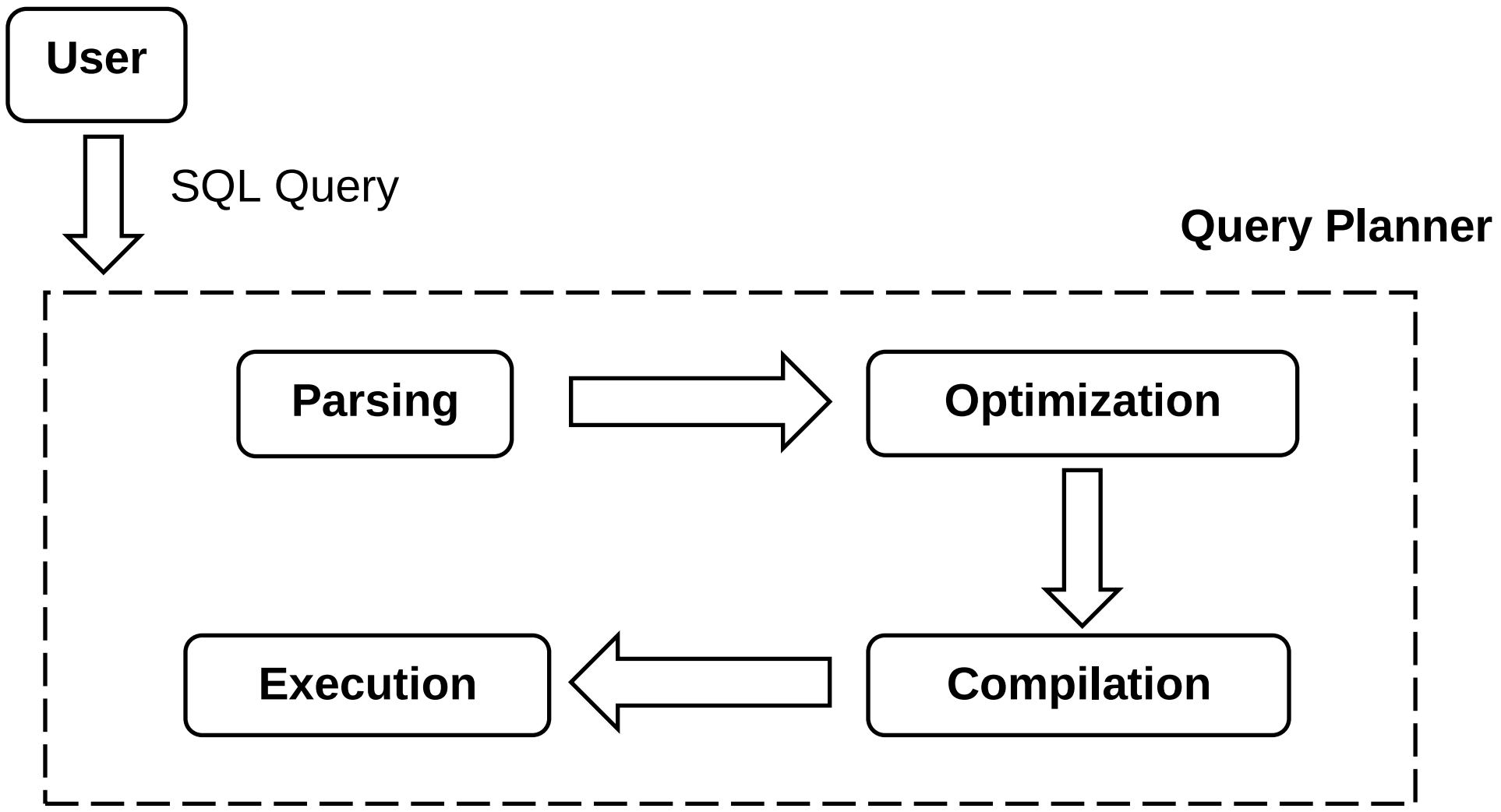

Figure 2.1: Architecture of a Query Planner

Let's look into the depth of how each of these components work and then look at an example of how these steps are implemented in GLADE. We will talk about query execution in detail in next chapter, when we explain how execution plans are converted to graph and waypoint files.

## 2.2 Query Parsing

When a SQL query is first submitted by the user, it is passed through a parser which translates the query into its equivalent relational algebra expression. The first step in query parsing is **Tokenization**. The query text is converted into its tokenize representation by the **query lexer**. We had already defined a set of rules which define how to scan a query and what action to perform against each token. This tokenize representation of each part of a query is in a form of a tree and is required by the parser for further **Semantics Validation**. The parser needs to check for the correct syntax of the query. This semantics validation involves determination of the validity of the query, like if all the tables are present in the database, the table attributes are valid and are present in the corresponding tables and other grammatical and structural validation. This step is necessary since we need to confirm that the format and all the variables of the query are correct.

In GLADE, we have used Lex/Yacc for tokenizing and finding the hierarchical structure for the query. We have predefined the tokens and the production rules which defines how to understand the input query language and what action to take for each statement the lexer has generated for parsing. Once the query passes all the parsing checks, it is then passed on to the query optimizer.

In order to understand how parser works, consider following query:

**SELECT DISTINCT** c_name, c_address, c_acctbal
**FROM** region, nation, customer, orders
**WHERE** o_custkey = c_custkey **AND** c_nationkey = n_nationkey
**AND** n_regionkey = r_regionkey **AND** r_regionkey < 5

Here, we have 4 relations/tables namely region, nation, customer and orders. The benchmark we have used throughout our project for testing was TPCH. The first step in parsing is to check if the structure of the query is correct or not. We see that the query has all the necessary clauses (SELECT, FROM, WHERE) and in the right order. We have defined these tokens in the parser dictionary already. The structure of a query in GLADE is break down into following:

- **SELECT Clause or Select List** (a linked list) with all the selection attributes and a linked list for functional operators, if any.
- **FROM Clause or Table List** (a linked list) with all the table names.

- **WHERE Clause or And List** (a tree) with all the joining conditions and the selectivity on attributes.

These three data structures are the output of a parser, which are then passed on to the next stage, query optimizer. In the case of this example, the parser creates each of these data structures while doing semantic validation at each step, checking for table names and attribute names for any grammatical or syntactical errors. If found any, the program terminates with an error.

In this case, our Select List will be a linked list with 3 attributes, c_name, c_address and c_acctbal, and it contains the information about any function being involved in the query or not, or in other words if the query is a simple query or an aggregate query:

**< Select List >** → c_name → c_address → c_acctbal
→ **DISTINCT**

There can be other functions too instead of DISTINCT, like a SUM or a GROUP_BY SUM or any other aggregate function. Aggregate functions are explained in details later in this section.

Table list will be a list of 4 tables:

**< Table List >** → region → nation → customer → orders

And List will be a tree where the root of each node corresponds to a single AND between two conditions and every condition can be either a comparison between two attributes in case of a JOIN, or it can be an attribute and a constant in case of a SELECT. The root of each condition corresponds to the operator and left and right nodes corresponds to the attributes.

Here is the tree constructed for this example:

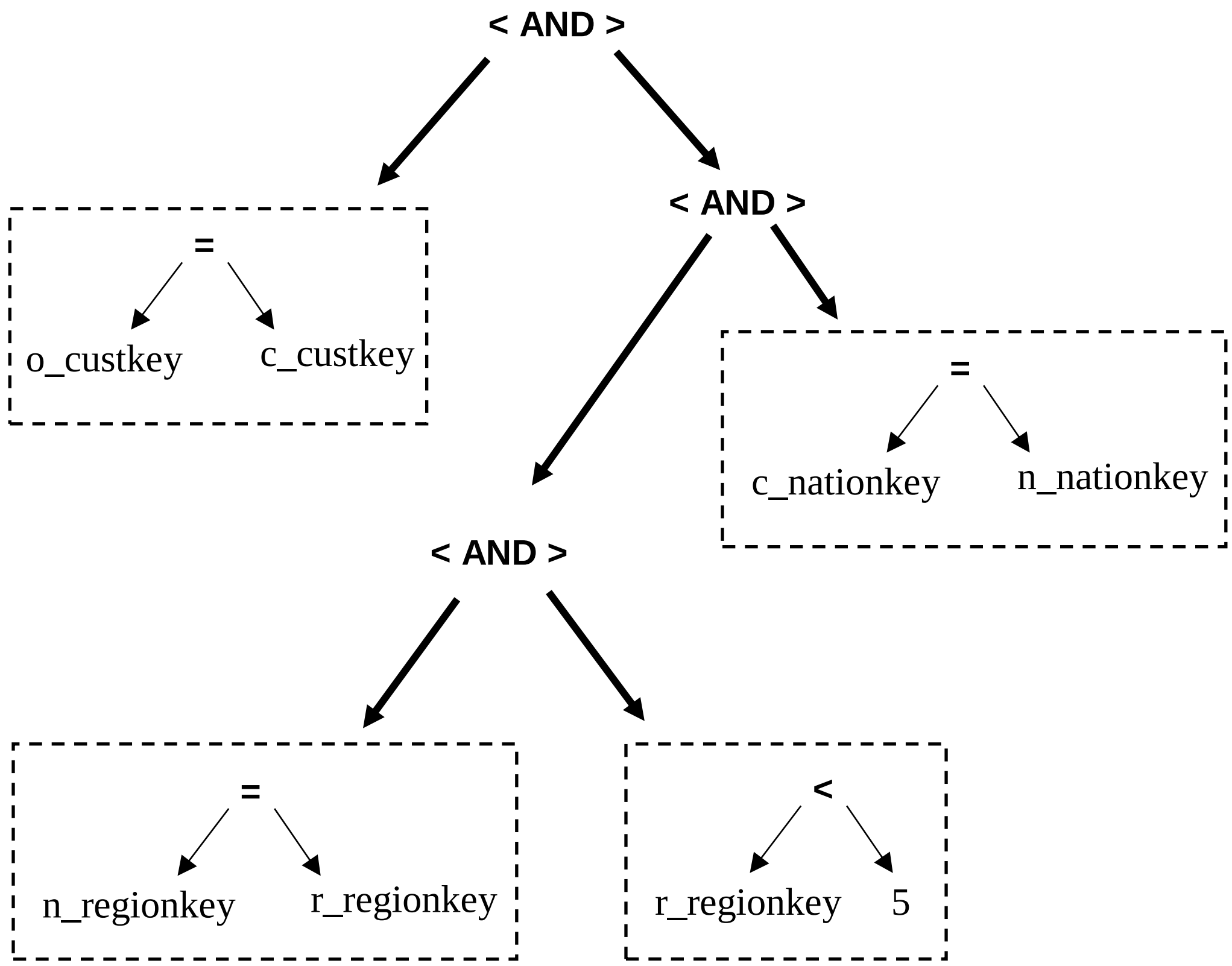

Figure 2.2: Parse tree of AND List condition in a query

## 2.3 Query Optimization

Query optimizer is the heart of the query planner which generates the optimal join plan to be eventually compiled and executed. Before we get into the optimizer, we need to understand the relational algebra operators, scan, select and join.

The most basic relational algebra operator is a **SCAN** operator, which can read the contents of an entire relation, say R. In our case, the whole relation is stored arbitrarily in the memory and it can be in any order, so in order to respond to any user query, we have to perform a full scan of the relation tuple by tuple. We call this as a table scan. There is a SCAN operator for every table in the query, so the total number of SCAN operators in any query are the total number of relations/tables to be scanned, and we can get this information from the WHERE clause of the query. A slight variation of this SCAN operator is a SORT-SCAN operator, which keeps on sorting the tuples read from the memory in order to maximize the efficiency at a later stage for other operators (such as SELECT<table name> or ORDER-BY operators).

The operator right on top of SCAN is a **SELECT<table name>** operator. This is an optional operator and depends on any selectivity being performed on any particular relation in the query e.g. scanning an entire table of customers but only picking those whose ID is less than 1000. This information of SELECT<table name> operator can be found in the WHERE clause of the query (the SELECT operator is different from SELECT clause of the query). The <table name> shows on which table is this selectivity being performed. For example in the example above, we have a selectivity on region, and so the operator would be *SELECTregion*.

The operator on top of SELECT (or SCAN) is a **JOIN** operator. This JOIN operator depends if there are any joins being performed between multiple relations in a query. If there are more than one relations in a query, then there is a JOIN operator corresponding to a join between two relations upon a certain attribute. This information can be found in the WHERE clause of the query too.

Other than these 3 operators, there are more operators which we will discuss more in detail in the next section. The main goal of query optimizer is to determine a realistic efficient join strategy of the operators for executing the query. There are 2 forms of query optimization:

- Rule Based Optimization
- Cost Based Optimization

### 2.3.1 Rule Based Optimization

Rule Based or Heuristic optimization technique comprise of generating the optimal join plan based on certain heuristic rules. The optimality of the join order is defined by the efficiency of the heuristic function. This method is beyond the scope of this paper since we worked on cost based query optimization in GLADE for generation of query plans which is explained below.

### 2.3.2 Cost Based Optimization Algorithm

On the other hand in Cost Based Optimization (CBO), each operator is assigned a ***cost*** which be any statistics collected from the database or any hardware parameters. We used the cardinality of each table to define the cost of the operator. This cardinality is determined using Flajolet-Martin algorithm (the detail of which is beyond the scope of this paper). At a bottom level, when defining the costs to each SCAN/SELECT operator, no optimization is required. Whereas at a higher level when we have multiple JOIN operators, we need to apply cost based optimization to get optimal join plans. This is the technique we used in our project and will be dicussed in the rest of the paper. The algorithms we used for CBO are defined under.

### 2.3.3 Algorithm

The dynamic program for a query using CBO algorithm is:

---

**Algorithm 1: Cost Based Optimizer Algorithm**

---

**Input**:

- Table names and total number of tables
- Selection predicates
- Join predicates

**Data structures:**

Map : map <table_name> ← tuple (size, cost, order)
where:

- table_name : a name that puts all the table names together, when there are more
- size : the cardinality, i.e., number of tuples, in table(s)
- cost : the sum of intermediate number of tuples in the tree (query optimization cost)
- order : join order with groupings of tables explicit

**Pre-processing:**

- Fill the map for every individual table name , say T

Map <(T)> = (size of the table from catalog, 0, (T))
// initial cost for each table is zero

- Push-down selections: update the size for each table based on the estimation of number of tuples following the application of selection predicates , if there are any

Map <(T)> = (size after push-down selection, 0, (T))

- Fill the map for every pair of two tables $T_1$ and $T_2$

Sort ascending $T_1$, $T_2$ (no need to have entries in Map for ($T_1$, $T_2$) and ($T_2$, $T_1$) because of commutativity)
Estimate cardinality of join using distinct values from Catalog and join predicates
Map <($T_1$,$T_2$)> = (Map <($T_1$)>.size * Map <($T_2$)>.size/selectivity, 0, ($T_1$, $T_2$))

- After we have all the pairings done, call the algorithm that generates all possible permutations for all {3, 4, 5 .. n} tables.

---

**Function Partition (N, Array[tables])**

```
if (Map <sort-asc(T1,T2,...,Tn)> is not empty) then
    // We already have an entry for this combination of tables
    return
end if

for i = 1 to N! do
     // Get (T1',T2',...,Tn'), the ith permutation of N tables.
    permutation(T1',T2',...,Tn');
    for j = 1 to N-1 do
        left = (T1',...,Tj')
        if (Map <sort-asc(left)> is empty) then
            Partition (j, left)
        end if

        right = (T(j+1)',...,Tn')
        if (Map <sort-asc(right)> is empty) then
           Partition (N-j, right)
        end if

        // at this point we are guaranteed to have entries in the
        // Map on left and right compute the cost for this tree
        // and store, if minimum; otherwise, discard

        cost = Map <sort-asc(left)>.cost + Map <sort-asc(right)>.cost
        if (j != 1) then
            cost += Map <sort-asc(left)>.size
        end if
        if (j != N-1) then
            cost += Map <sort-asc(right)>.size
        end if
        if (cost < min_cost) then
        // estimate cardinality of join using distinct values
            size = Map <(left)>.size * Map <(right)>.size/selectivity
            order = Map <sort-asc(left)>.order + Map <sort-asc(right)>.order
            min_cost = cost
        end if
     end for
end for
Map <sort-asc(T1,T2,...,Tn)> = (size, min_cost, order)
```

**Function Permutation(array [$T_1$',$T_2$',...,$T_n$'])**

```
// This algorithm generates the next permutation lexicographically
// provided a given permutation.

// Find the largest index k such that a[k] < a[k + 1]

k← -1
l ← 0;
for i = 0 through total number of tables (n) do
    if array[i] < array[i+1] then
        k = i;
    end if
end for

// If no such index exists, the permutation is the last permutation.
if k = -1 then
    return false;
end if

// Find the largest index l greater than k such that a[k] < a[l]
for i = k+1 through total number of tables (n) do
    if array[k] < array[i] then
        l = i;
    end if
end for

// Swap the value of a[k] with that of a[l]
swap(array[k], array[l]);

// Reverse the sequence from array[k + 1] up to and including the final
// element array[n]
count ← 1
temporaryIndex ← k+1;

for i = 0 through (n- k-1)/2 do
    swap(array[ temporaryIndex ], array[ n-(count++) ]);
end for
return true;
```

## 2.3.4 Example Explanation

In order to understand the algorithm, consider the same example we saw in the previous section:

**SELECT DISTINCT** c_name, c_address, c_acctbal
**FROM** region, nation, customer, orders
**WHERE** o_custkey = c_custkey **AND** c_nationkey = n_nationkey
**AND** n_regionkey = r_regionkey **AND** r_regionkey < 5

For simplicity, we will use the following table encoding for relations we have:

| | |
|---|---|
| *region* | *0* |
| *nation* | *1* |
| *customer* | *2* |
| *orders* | *3* |

### 2.3.4.1 SCAN

In the preprocessing stage of the algorithm, the first step is to initialize the data structure. This is the implementation of a SCAN operator, which scans tuples from each table. The cost of all the operators is set to zero, the size of each table is initialized as the cardinality of the tables from TPCH database. This initial size shows that each table has to be scanned completely unless there are any further constrains on it. Since there are no relations joined yet, so the join order or simply the order is set to just the name of the relation itself. Here is how the output after the first step looks like:

*0 → (size: 5, cost: 0, order: 0)*

*1 → (size: 25, cost: 0, order: 1)*

*2 → (size: 150000, cost: 0, order: 2)*

*3 → (size: 1.5 x* $10^6$*, cost: 0, order: 3)*

### 2.3.4.2 Push-down Selection

The next step is to implement a SELECT operator, if any. In other words we need to update the size of each relation based on any selectivity in the query. This is called push down selection. Push down selections basically help in reducing the size of a relation and is an efficient approach in query optimization which pushes the selections down in the tree as far as we can. This is one of the most powerful tool of query optimizer since we apply the selections before the JOIN operator (that's the furthest it can be pushed down). By doing so guarantees that when the JOIN operator is applied, we don't get an inundation of tuples, hence reducing the size of the relation by a large factor at an early stage just before applying a giant operator (JOIN).

In order to estimate the size reduced by a SELECTregion operator, there are following possible cases, when:

i. An attribute is equated to a constant using equality operator, in which case the size is reduced by a factor of total number of distinct elements for that attribute.

$$Size\ after\ selectivity = \frac{Size\ before\ selectivity}{Number\ of\ distinct\ elements}$$

If T(R) represents size of a relation R, and if we assume that we had the case r_regionkey = 5 instead in the example above, then size after selectivity would be given by:

$$T(\sigma_{\text{r_regionkey = 5}}\ (region)) = \frac{T(region)}{V(region, r_region)}$$

where $\sigma_{\text{r_regionkey = 5}}$ is the selectivity on the attribute and V(X, Y) shows the number of distinct elements in relation X over attribute Y.

ii. An attribute is compared to a constant using greater or less than operator. [5] and [6] explains how more problematic it gets when the selection involves comparison, just as in case of this example. Most of the databases stores data in columnar format and the data follows a highly skewed distribution, for example *Zipfian Distribution* or *Normal Distribution*. On average, in every comparison operator, we can assume that half of tuples can satisfy the comparison operator. So T(R)/2 would be a good estimate for for the relation. But this is not a good heuristic in every scenario. Mostly, more than half of tuples lie below or above the given constant just as in the case of negatively skewed or positively skewed distributions. So a better heuristic would be to estimate that the selectivity would return somewhere around one third of the total tuples and not half. In case of our example, the size of the selectivity would be given by:

$$T(\sigma_{\text{r_regionkey} < 5}(region)) = \frac{T(region)}{3}$$

The cost would, however, remain zero, since no relations are joined yet. So in the case of our example, after applying push-down selections, we get the following output:

*0 → (size: 1.66667, cost: 0, order: 0)*

*1 → (size: 25, cost: 0, order: 1)*

*2 → (size: 150000, cost: 0, order: 2)*

*3 → (size: 1.5 x $10^6$, cost: 0, order: 3)*

iii. An attribute is not equal to a constant. This is rather a very rare case and is very rarely used in the queries. A selection of type r_regionkey ≠ 5 would essentially return almost the same amount of tuples. Hence we estimate in this case as:

$$T(\sigma_{\text{r_regionkey} \neq 5}(region)) = T(region)$$

Alternatively, we can also use the fact that we have one tuple less than the tuple we are comparing the relation with, so in that case we will have:

$$T(\sigma_{\text{r_regionkey} \neq 5}\,(region)) = \frac{T(region) * (V(region, r_region)\ - 1)}{V(region, r_region)}$$

which is slightly less than the total number of tuples, but in the case of extremely large database, this would not matter much. So we estimate that in the case of inequality, the selectivity factor is 1.

### 2.3.4.3 JOIN

The next step in the algorithm is to make pairs of all the relations. For any two relations, $T_1$ and $T_2$, there can be only one pair of ($T_1$,$T_2$) because of the commutative property, and we would not consider ($T_2$, $T_1$). We wish to compute the joins of all the possible relations and calculate the resulting sizes of each joins. For our testings, we only considered natural joins of two relations, with only the cases when one attribute is equal to the other attribute. Before we go into the results we get by joining all the relations in the case of our example, let us look at a case of joining simple relations and how they are joined in different scenarios. Consider two relations $R(X, Y) \bowtie S(Y, Z)$ with X,Y and Z as the attributes. The problem is we are not sure about Y-values in both R and S and how are they related, for example:

- Both relations can have a disjoint sets of the joining attribute, in which case we cannot join the two relations and the join results in empty set:

$$T(R \bowtie S) = 0$$

- The tuples of both relations can have same value for the joining attribute, in which case the join is the cartesian product of number of tuples of both relations:

$$T(R \bowtie S) = T(R) * T(S)$$

In order to focus on some of the most common situations, we need to understand following two rules:

- **Encompassing attributes**: Consider an attribute Y present is various relations and consider it being present in both R and S, the two relations we have, such that the number of tuples with distinct values of Y in R is less than the number of tuples with distinct values of Y in S, i.e. $V(R,Y) < V(S,Y)$, then every Y-value of R will be a Y-value of S.
- **Preserving attributes**: Consider a third attribute P which is not a join attribute and which is present is R but not in S and we are to join R and S. Then the attribute P will not loose its values after the join is encountered. In other words:

$$V(R \bowtie S, A) = V(R, A)$$

Following these rules, we can now estimate the size of the join between R and S. Assume that *r'* is a tuple in R and *s'* is another tuple in S. We need to find the probability that both *r'*and *s'* are a part of attribute Y. Suppose that $V(R,Y) \geq V(S,Y)$. Then all the Y-values of S are surely present in the Y values of R under the light of encompassing attributes rule. Hence the probability that both *r'*and *s'* has the same Y-value is $\frac{1}{V(R,Y)}$ . On the other hand, if $V(R,Y) < V(S,Y)$, following the same criteria as above, we will get the probability that both *r'*and *s'* has the same Y-value is $\frac{1}{V(S,Y)}$ . Hence in general, we estimate that the probability of Y-value when joining two relations is:

$$P(Y \mid Y \in R \,\&\, Y \in S) = \frac{1}{max(V(R,Y), V(S,Y))}$$

Hence

$$T(R \bowtie S) = T(R) * T(S) * \frac{1}{max(V(R,Y), V(S,Y))}$$

which says that the estimated number of tuples after a JOIN is applied between two relations/tables is equal to the Cartesian product of the number of tuples from each relation (after applying push down selections) times the probability that each such pair shares the same attribute.

Getting back to the example, the algorithm computes the joins by picking each table and looping it over the rest of the tables and estimating the size of each JOIN using the strategy discussed above. Consider the first case *01*, where *0* is *region* and *1* is *nation*, and we have a join between them on the attribute *regionkey* in the query. From the data we have uptil now:

$T(region) = 1.66667$

$T(nation) = 25$

$V(region, r_regionkey) = 5$

$V(nation, n_regionkey) = 5$

$$T(region \bowtie nation) = 1.66667 * 25 * \frac{1}{max(5{,}5)} = 8.33333$$

The join order of this pair is going to be *01* and the cost of this pair is still zero. This is because of the fact that if we were to stop the algorithm at the first stage of the JOIN, then that can be the case only when there are only 2 tables to be joined. In that case, the algorithm will only generate one pair and that one pair is going to be the optimal join order for that query. So we do not worry about cost at this stage. On the other hand, if the algorithm generates more than one pair of joins, then we have more than 2 tables, so the initial join cost for the individual pairs does not matter. It is the cost of the next level of JOIN with more than 2 tables involved.

Following the same pattern and generating all the pairings, we would get the following pairs:

| **Relation** | **Size** | **Cost** | **Join Order** |
|---:|---:|---:|---:|
| 0 | 1.66667 | 0 | 0 |
| 01 | 8.33333 | 0 | 01 |
| 02 | 250000 | 0 | 02 |
| 03 | $2.5 \times 10^6$ | 0 | 03 |
| 1 | 25 | 0 | 1 |
| 12 | 150000 | 0 | 12 |
| 13 | $3.75 \times 10^7$ | 0 | 13 |
| 2 | 150000 | 0 | 2 |
| 23 | $1.5 \times 10^6$ | 0 | 23 |
| 3 | $1.5 \times 10^6$ | 0 | 3 |

After generating all the possible pairs, we call the core of the algorithm, which generates all the possible permutations starting from 3 tables and going to total number of tables in a given query. For each permutation a minimum cost is calculated along with the size of JOIN. The algorithm calculates a left sub-string for the left node of the JOIN and a right for the right node of JOIN. *The cost of the JOIN is equal to the sum of sizes of left and right nodes*. The size of each node is calculated in the same way as it was calculated for the pairs in the previous step. At the end of each iteration, the minimum cost for a specific JOIN is stored for that order.

**Lexicographical Permutation Generator Algorithm**: We used the simple algorithm to generate the permutations in a lexicographical order. At each level, all the possible permutations of the tables are generated in a lexicographical order. Consider the first stage, when we are trying to find all the permutaions for *012*. The total number of possible permutations for *012* are 3! = 6. The algorithm starts with all the tables in ascending order in a string and ends with the last permutation, when all the tables are in descending order. According to the algorithm:

- Find the largest position, say $L$ in the string array, such that

  $$Array[L] < Array[L+1]$$

  If there is no such index found at the end of the loop, then this is the case when we have the string in descending order. Hence we have reached the end of our permutations for this level and the algorithm terminates.
- If $L$ exists, we need to find a second largest index, $L'$ after $L$, such that

  $$Array[L] < Array[L']$$

- Swap the values of both $Array[L]$ and $Array[L']$.
- Start from $Array[L+1]$ to $Array[Size\ of\ Array]$ and reverse the whole sequence.

In the case of *012*, the algorithm will generate the permutations "*012, 021, 102, 120, 201, 210*". Each single permutation is associated with a unique cost calculated by the algorithm.

**Greediness of the Algorithm:** The approach the algorithm aplies for finding the optimal order at every level is ***greedy*** because we consider one decision at one time about the order of joins and never backtrack or consider altering the decision we already made. The ***greediness*** is based on the fact that we need to keep the statespace as small as possible. Hence we keep discarding join orders with higher cost and only consider the one with the least cost and move to the next level.

Consider a sub-string *012*. In the first step of the first iteration, the algorithm will first pick *0* as a left node and *12* as the right node from the list of previously observed nodes, and calculate a cost of joining *0* with *12*. Then in the next step, it will pick up *01* as a left node and *2* as a right node and again calculate the cost for the order *012*. The process repeats for all the permutations for 012 and in the end the cost of which ever join order is least will be stored in the map along with its respective JOIN size and along with the JOIN order. The JOIN order is calculated using parenthesis, which indicates a JOIN, for example, *(01)2* indicates a JOIN of *2* with a JOIN of *01*.

The structure of JOIN node in the form of tree looks like:

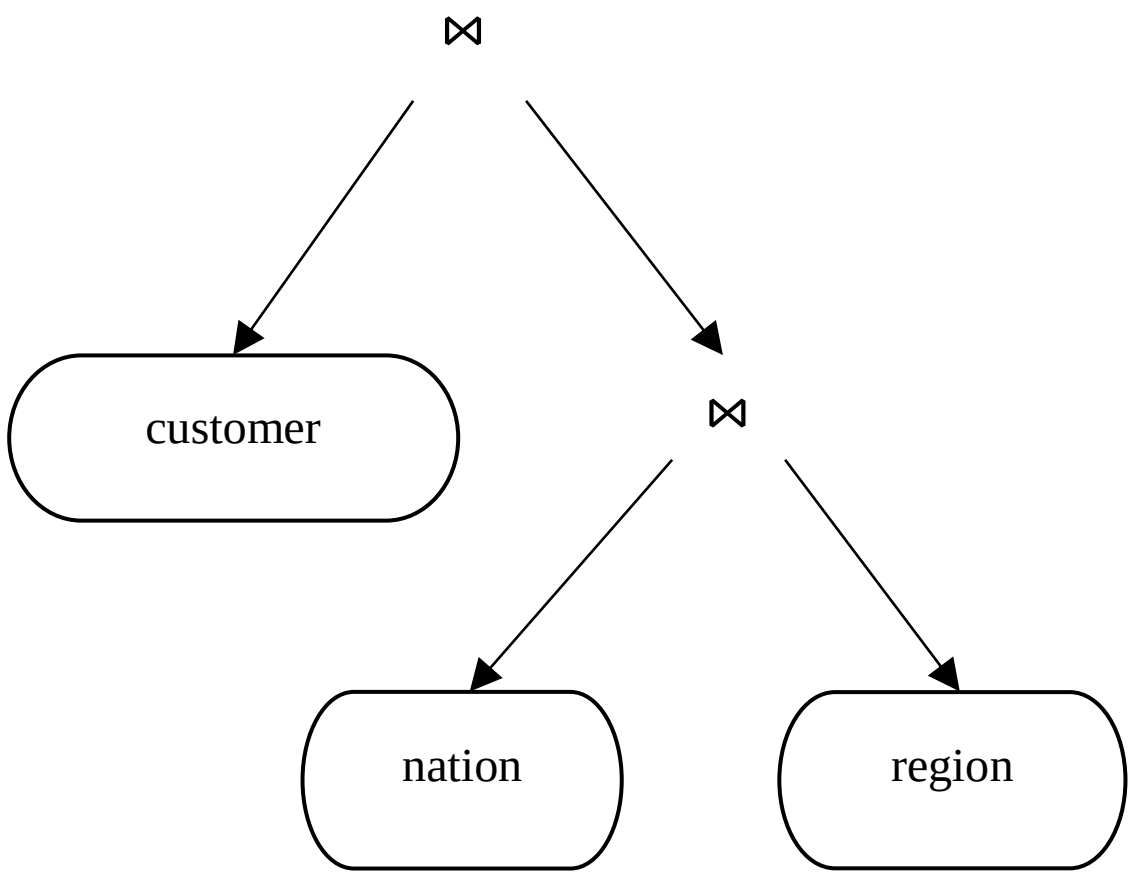

Figure 2.3: JOIN order of an intermediate stage in CBO algorithm

After a single iteration is completed, the final join order is updated in the map so it can be further used as a sub-node for the following iterations. The total number of iterations depends on the total number of tables, and at each level, total number of permutations depends on the permutation algorithm.

For our example, the algorithm populates the following complete map with each permutation having the optimum least cost and size along with the join order. We then pick up the complete joining order which includes all the tables *0123* and generate an execution tree in the next stage.

| **Relation** | **Size** | **Cost** | **Order** |
|---|---|---|---|
| 0 | 1.66667 | 0 | 0 |
| 01 | 8.33333 | 0 | 01 |
| 012 | 50000 | 8.33333 | (01)2 |
| **0123** | **500000** | **50008.3** | **((01)2)3** |
| 013 | $1.25 \times 10^7$ | 8.33333 | (01)3 |
| 02 | 250000 | 0 | 02 |
| 023 | $2.5 \times 10^6$ | 250000 | (02)3 |
| 03 | $2.5 \times 10^6$ | 0 | 03 |
| 1 | 25 | 0 | 1 |
| 12 | 150000 | 0 | 12 |
| 123 | $1.5 \times 10^6$ | 150000 | (12)3 |
| 13 | $3.75 \times 10^7$ | 0 | 13 |
| 2 | 150000, | 0 | 2 |
| 23 | $1.5 \times 10^6$ | 0 | 23 |
| 3 | $1.5 \times 10^6$ | 0 | 3 |

The final join order containing all tables is guaranteed to have the least join cost. As a rule of thumb, any relation with fewer number of tuples becomes the right node in the tree. The final join order tree for this query, represented by *((01)2)3* is shown below:

Figure 2.4: Complete JOIN order of CBO algorithm

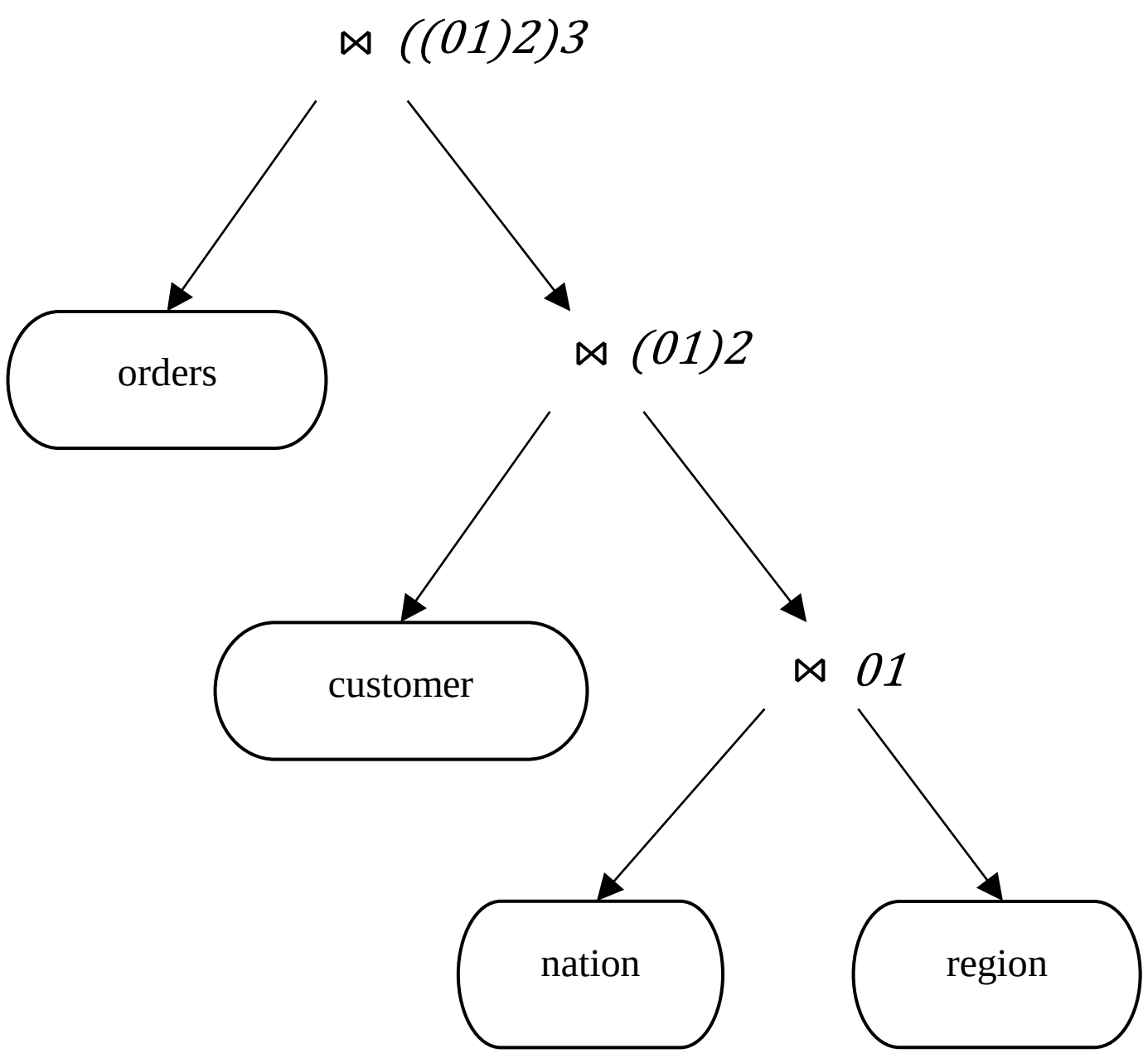

#### 2.3.4.4 Complexity

The time complexity of the whole CBO algorithm described in 2.1.1.3 is solely determined by the complexity of the lexicographical order algorithm. The worst case time complexity of lexicographical algorithm is $O(x^2 * x!)$, where $x$ is the size of the string array.

For total of *n* number of relations/tables to be scaned, where *n > 2,* the overall time complexity of the algorithm turns out to be:

$$\sum_{i=3}^{n} O(i^2 * i!) \approx \boldsymbol{O(n^2 * n!)}$$

## 2.4 Query Compilation

After the initial optimal join order tree is generated by the Query Optimizer, the next step in query planning is to constrct a complete query execution tree by implementing the rest of relational algebra operators in the query (if any). The first step in that is to generate the execution tree following the join order created by the optimizer.

A single execution node in our project consists of a left pointer to the left node, a right pointer to the right node, a parent pointer to the parent node and the name of the node (SCAN, JOIN etc). A dynamic program of the algorithm is given below:

### 2.4.1 Algorithm

**Algorithm 2: Construct Execution Tree Algorithm**

```
// This algorithm constructs the execution tree provided a given optimization tree.
Input: Root pointer to optimization tree: root
Output: Query execution tree node: qetn

QueryExecutionTreeNode constructTree(QueryOptimizationTreeNode):
    if leftChild is null and rightChild is null then
        // You have reached the end of tree, create a new execution node
        qetn ← new query execution tree node
        qetn.leftChild is null
        qetn.rightChild is null

        // Find the name of this node in the list of SELECT and SCAN(s)
        if select.isThere(root) then
                // if it is a SELECT node, a new qetn is to be inserted between
                // SCAN and the parent node
                qetn_new ← new query execution tree node
                qetn_new.name = "SELECT" + root.name
                qetn_new.parent = qetn
        end if
        else
                qetn.name = root.name
        end else
        return qetn
    end if

    else // You have reached a JOIN node
        // Find the left and the right execution tree nodes by recursively calling
        // the function
        qetn left = constructTree(root.leftChild)
        qetn right = constructTree(root.rightChild)

        qetn_new ← new query execution tree node
        qetn_new.leftChild is left and qetn_new.leftChild.parent is qetn_new
        qetn_new.rightChild is right and qetn_new.rightChild.parent is qetn_new
        qetn_new.name = "JOIN" + i^th number of JOIN node in the tree
        return qetn_new
    end else
```

If we consider the same example as we considered in the last section, the algorithm generates the the following execution tree:

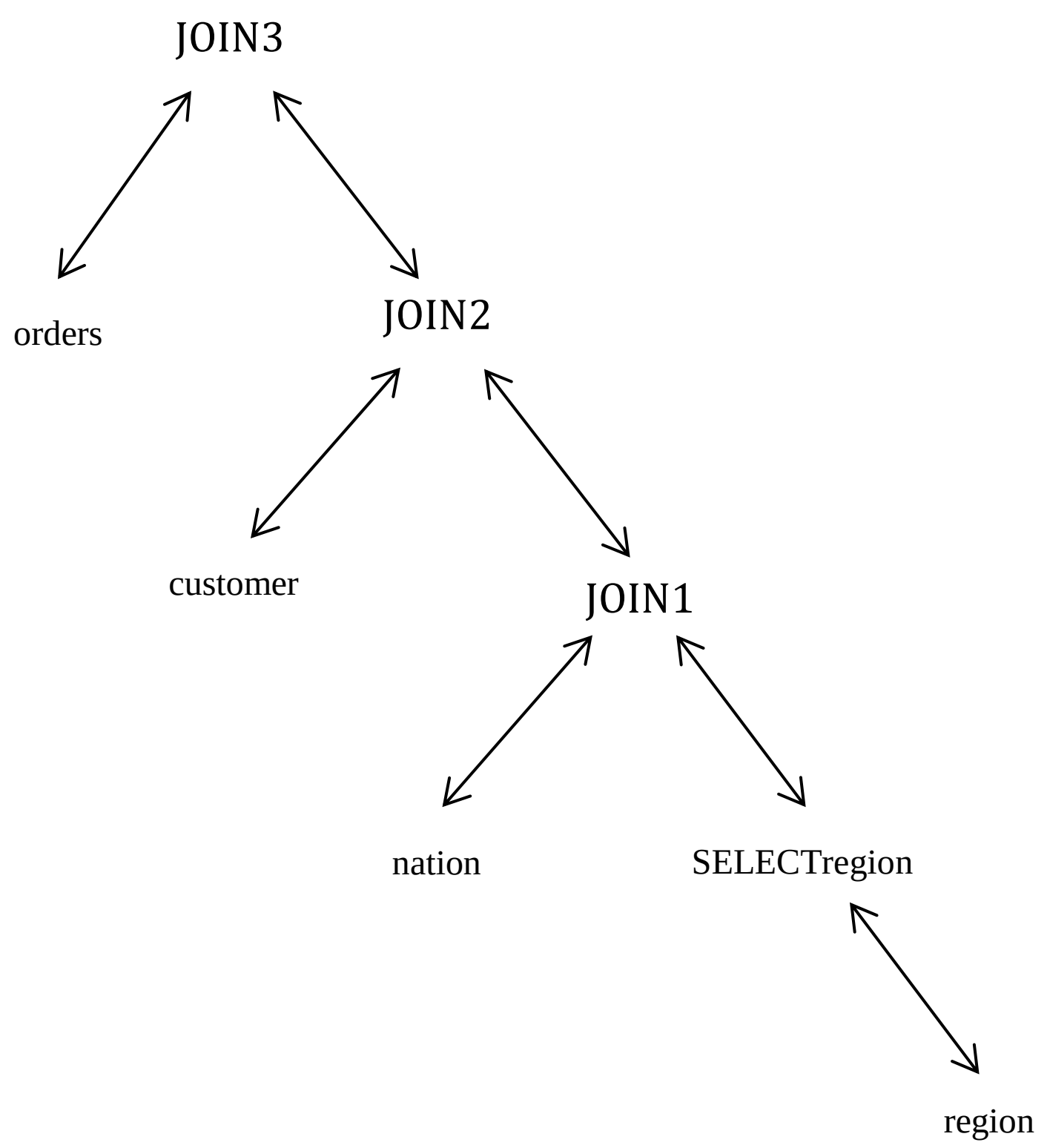

Figure 2.5: Execution Tree

After query execution tree generation, remaining operators are being added on top of the root of this tree. There can be two different types of queries that we considered:

- Simple queries involving Select-From-Where clause only
- Aggregate queries involving aggregate functions or user defined aggregates (generalized linear aggregates in GLADE terminology)

## 2.4.2 Simple Queries

In case of a simple SELECT-FROM-WHERE queries, there are two cases. We can either have just a projection or finding distinct number of elements in the query.

**PROJECT**: The PROJECT operator is performed at the top of the query execution tree. It performs the projection on each tuple coming out of the execution tree based on the attribute in the SELECT clause of the query. So every simple query has a PROJECT at the top of the tree depending on how many attributes we have in the SELECT clause. The way how tuples are projected depends on the algorithm we use for joining the tables.

In case of our sample example, there are 3 attribute, *c_name, c_address, c_acctbal*. The PROJECT operator gives the information about which attributes to keep from all the joining attributes coming down from the entire execution tree. For this example, only *c_name, c_address, c_acctbal* are kept from all the tuples.

**DISTINCT:** The DISTINCT operator also known as duplicate elimination operator, if any, is performed on top of PROJECT. The duplicate elimination or DISTINCT on the tuple level can be done by performing either hashing or sorting at the JOIN level because it is easy to eliminate duplicate and keep distinct elements if the whole state space is sorted. As the name says, the job of DISTINCT operator is to remove any duplicate entries after the projection is being performed, on only the attributes that are left after the PROJECT operator.

In case of our example query, we have a DISTINCT operator involved in it. So query will first project all the records based on the 3 attributes *c_name, c_address, c_acctbal*, and then it performs a DISTINCT operator to discard all the duplicate tuples. A complete query execution tree for this query will look like:

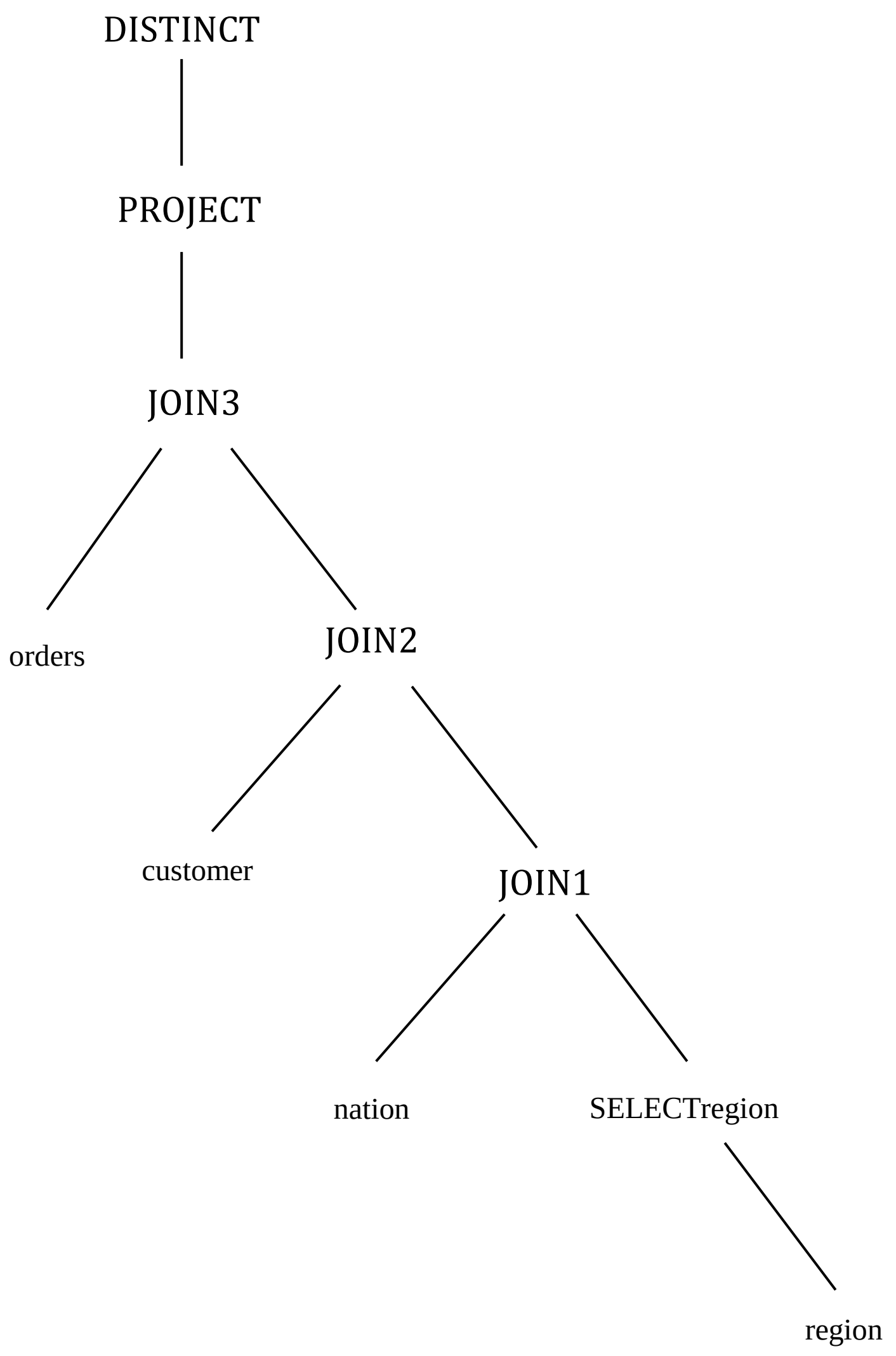

Figrue 2.6: Complete Query Execution Tree for the query:

**SELECT DISTINCT** c_name, c_address, c_acctbal
**FROM** region, nation, customer, orders
**WHERE** o_custkey = c_custkey **AND** c_nationkey = n_nationkey
**AND** n_regionkey = r_regionkey **AND** r_regionkey < 5

### 2.4.3 Aggregate Queries

Queries involving an aggregate function are more complicated than a simple query. There can be many user defined aggregate functions (UDAF) depending upon the user need. In GLADE, we call these UDAF as GLA and they are discussed in Chapter 4. For the testing purpose, we only considered 2 default aggregate functions, which will be explained here, the SUM and GROUP BY aggregate functions. The example we are considering uptil now does not involve any aggregate function, but we are going to explain different example here in order to explain how aggregate function works and how the execution tree changes with the addition of these functions. Consider an aggregate version of the query we have been discussing:

**SELECT SUM** (c_acctbal), c_name
**FROM** region, nation, customer, orders
**WHERE** o_custkey = c_custkey **AND** c_nationkey = n_nationkey
**AND** n_regionkey = r_regionkey **AND** r_regionkey < 5
**GROUP BY** c_name

**SUM**: For the queries only involving SUM, a SUM node is added on top of the optimization tree. It keeps the running sum of the attribute in the argument of it. In case of above example, the SUM will keep a running sum of *c_acctbal* from all the tuples that the bottom node is generating.

**GROUP BY**: A GROUP BY operator is actually a hybrid of DISTINCT and a SUM and it is always followed by SUM. So if there is a GROUP BY node, there is a SUM node already infused in it, so we don’t add a separate SUM node. Hence there can be either a GROUP BY or a SUM node in the execution tree, but not both together.

The way how GROUP BY works is that it creats groups based on the attribute in the GROUP BY clause and then perform the aggregate function on each group based on the attribute in the SUM clause (the aggregate function can be a sum, count, min or max from the grouping attribute, but we are only considering ‘sum’ for our testing purposes). In case of this example, it will keep a running sum of *c_acctbal* and create groups according the *c_name*. A complete query execution tree for this query will look like:

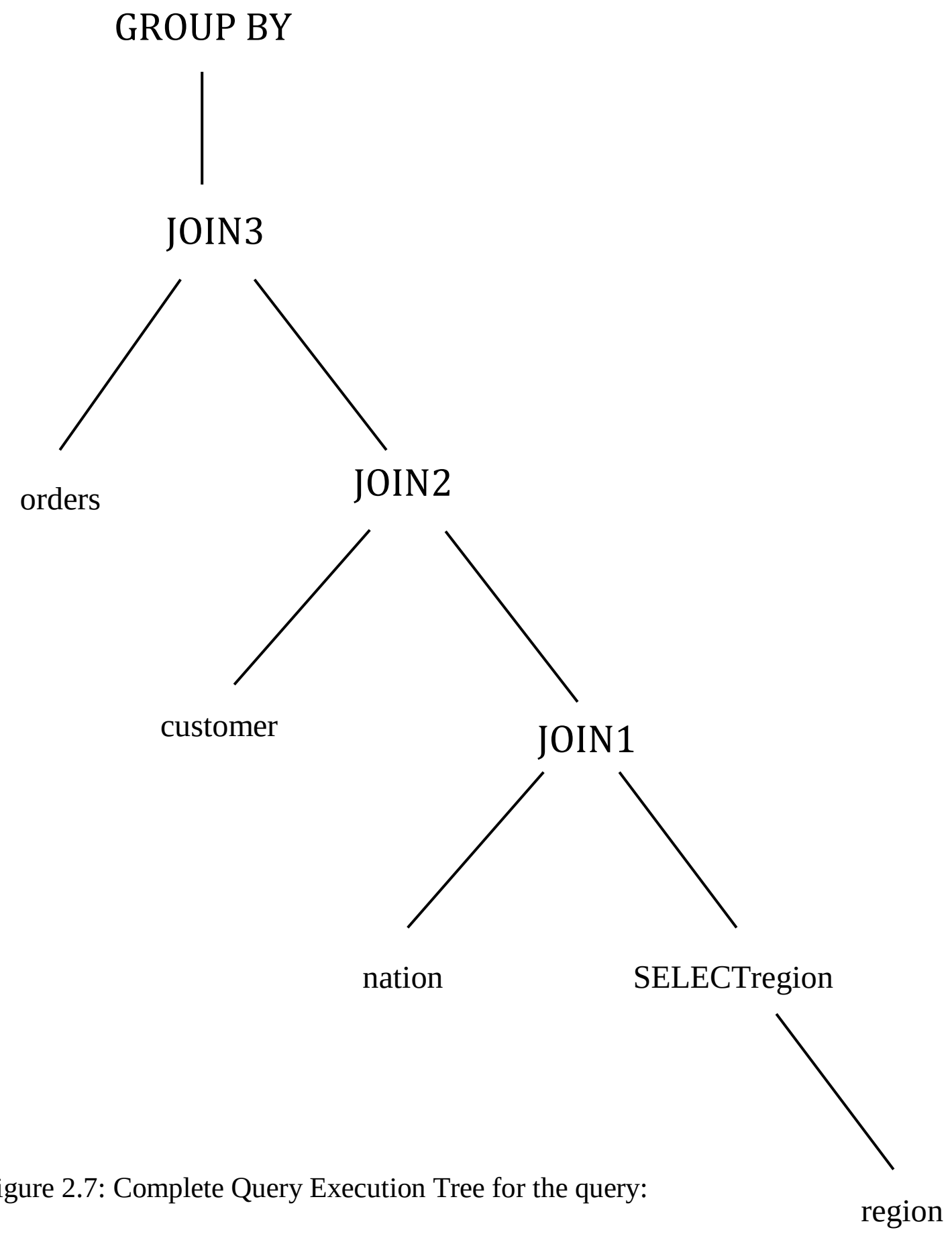

Figure 2.7: Complete Query Execution Tree for the query:

**SELECT SUM**(c_acctbal), c_name

**FROM** region, nation, customer, orders

**WHERE** o_custkey = c_custkey **AND** c_nationkey = n_nationkey

**AND** n_regionkey = r_regionkey **AND** r_regionkey < 5

**GROUP BY** c_name

# Chapter 3

# Query Execution

## 3.1 Introduction

We have seen in the previous chapter that how optimization plans are generated and are converted to execution plans containing execution nodes. By this time, we are sure about the validity of the query because it has been scanned, parsed, planned and compiled. The query execution tree nodes helps to generate 2 very important files, called Graph and Waypoint files, which guides GLADE how to construct the network of nodes in order to extract the data and return to the user. We will see in more detail about how GLADE architecture looks like and what are main jobs of some of the components in GLADE, but for now, we will explain the process at a high level. The main purpose of query executor is to initialize the number of nodes needed, organize them according to the execution plan and log any run time errors if it encounters. The graph file has all the information about what nodes are required to initialize and what are the leaf nodes, internal nodes and the root node. The waypoint file on the other hand defines the structure of nodes and what processing is to be done at each node at the tuple level. They are explained below in detail.

## 3.2 Graph File

For generation of graph and waypoint files, we have only worked with simple SELECT-FROM-WHERE queries. Aggregate queries are left for future work. Consider the simple version of the same query we had discussed in the previous section:

**SELECT** c_name, c_address, c_acctbal
**FROM** region, nation, customer, orders
**WHERE** o_custkey = c_custkey **AND** c_nationkey = n_nationkey
    **AND** n_regionkey = r_regionkey **AND** r_regionkey < 5

The graph file generated for this query is shown below. We will learn in the next chapter that GLADE consists of 2 main type of nodes, a coordinator node and a worker node. The job of coordinator is to assign and define worker nodes which works on tuple level to join data according to the query execution plan. The graph file indicates that there are 4 tables to be scanned and we need 5 worker nodes for this. All the tables to be scanned are at leaf nodes and the 5 worker nodes are the intermediate node, out of which one is going to be the root node which will pass the data back to coordinator, which in turn returns it to the user. It defines the links between any nodes in the same way as it was defined in the execution tree we saw in the previous chapter. The PRINT node is the root node and always stays at the top, whose job is to terminate the process after final tuples are merged and then returns the results back to the user. How each of the node works and how the code is executed on each node is explained in the next section.

**4 TableScans**
customer
nation
orders
region

**5 WayPoints**
PRINT
JOIN1
JOIN2
JOIN3
SELECTregion

**1 Queries**
Q1

**4 Leaves**
customer
1 Queries
    Q1  JOIN3
0 TerminalLinks
1 RegularLinks
    JOIN2

nation
1 Queries
    Q1  JOIN2
0 TerminalLinks
1 RegularLinks
    JOIN1

orders
1 Queries
    Q1  PRINT
0 TerminalLinks
1 RegularLinks
    JOIN3

1

region
1 Queries
    Q1  JOIN1
0 TerminalLinks
1 RegularLinks
    SELECTregion

**5 Nodes**
PRINT
0 TerminalLinks
0 RegularLinks

JOIN1
1 TerminalLinks
    Q1 JOIN2
0 RegularLinks

JOIN2
1 TerminalLinks
    Q1 JOIN3
0 RegularLinks

JOIN3
1 TerminalLinks
    Q1 PRINT
0 RegularLinks

SELECTregion
1 TerminalLinks
    Q1 JOIN1
0 RegularLinks

2

## 3.3 Waypoint File

The waypoint generated for the same query is shown below.

```
4 TableScans

orders 1
Q1 PRINT 1 o_custkey 0

customer 1
Q1 JOIN3 5 c_custkey c_name c_address c_nationkey c_acctbal 0

nation 1
Q1 JOIN2 2 n_nationkey n_regionkey 0

region 1
Q1 JOIN1 1 r_regionkey 0

5 WayPoints

PRINT print 1
Q1 printList (val(c_name), val(c_address), val(c_acctbal))$

JOIN3 join 2
lefthash string o_custkey$
Q1 join (), ( (c_custkey),(c_name, c_address, c_acctbal))$

JOIN2 join 2
lefthash string c_nationkey$
Q1 join (c_custkey, c_name, c_address, c_acctbal), ( (n_nationkey),())$

JOIN1 join 2
lefthash string n_regionkey$
Q1 join (n_nationkey), ( (r_regionkey),())$

SELECTregion selection 2
Drop string $
Q1 selection ((val(r_regionkey) < 5))$
```

The waypoint file describes how a query is managed and how the different types of computations are assigned, like selection, join or printing (output to the user). The nodes of the waypoints which were defined in the previous step are now initialized with the code that needed to be executed in order to start getting chunks of data from the storage manager. The worker nodes does not execute any query processing job by themselves. The work is actually being done on the threads, which are initialized by the workers from a pool of free threads. Once the setup is done, the storage manager starts producing chunks which are routed towards the workers according to the execution plan. The execution plan at this point (the waypoint file) is loaded at each worker and each worker knows its job that what tuples are needed to be selected, what tuples are needed to be join if there is a join, and what tuples are needed to be pruned at this stage because the query does not need them at any stage after this. This all information resides in the waypoint file.

At a high level, how this works is simple. All the threads are initialized using the Init routine (we will discuss these routines in detail in next chapter). The table scaners tells us which attributes we will be looking for in each table. Once everything is initialized, the chunks start to accumulate. This is done in the Accumulate routine. Once each worker is done accumulating chunks and its job is done, it passes the data to its parent node. At this point, two cases can happen. If this worker is a leaf node, its only job is to serialize the data and send to its parent node. If its an internal node, it first receives the data coming from the underlying nodes, deserializes the data, merge the data with the data it contains (this is done in the Merge routine) and finally serializes the data and pass it to its parent node. During merge phase, the join waypoints have the information about which attributes from the left and right side will be arriving, which attributes from this node will we use to join and eventually which attributes we need to prune and which attributes we need to pass to the parent node. In the end, when all the merging is finalized, the PRINT node is executed which performs any final merging and then terminate the process by calling the Terminate routine and passes the results back to the user.

# Chapter 4

# GLADE Architecture

GLADE is short for Generalized Linear Aggregate (GLA) Distributed Engine, which is a scalable and distributed system for analyzing big data. This chapter put some light on the GLADE [7], [8] architecture in general and how query execution is implemented in GLADE.

## 4.1 Architecture

GLADE is scalable distributed system for the execution of analytical data by providing an engine to execute user defined aggregate functions at multiple levels inside a single node. The system is responsible for generating the execution tree and moving data within the network from storage to client. The data resides in the memory in a column-oriented format and the storage system itself is a relational execution engine which is derived from DataPath [2]. DataPath works as a shared-memory single node processing system, while GLADE on the other hand has a shared-nothing cluster architecture and is a multi-node distributed system. The architecture of GLADE is shown below.

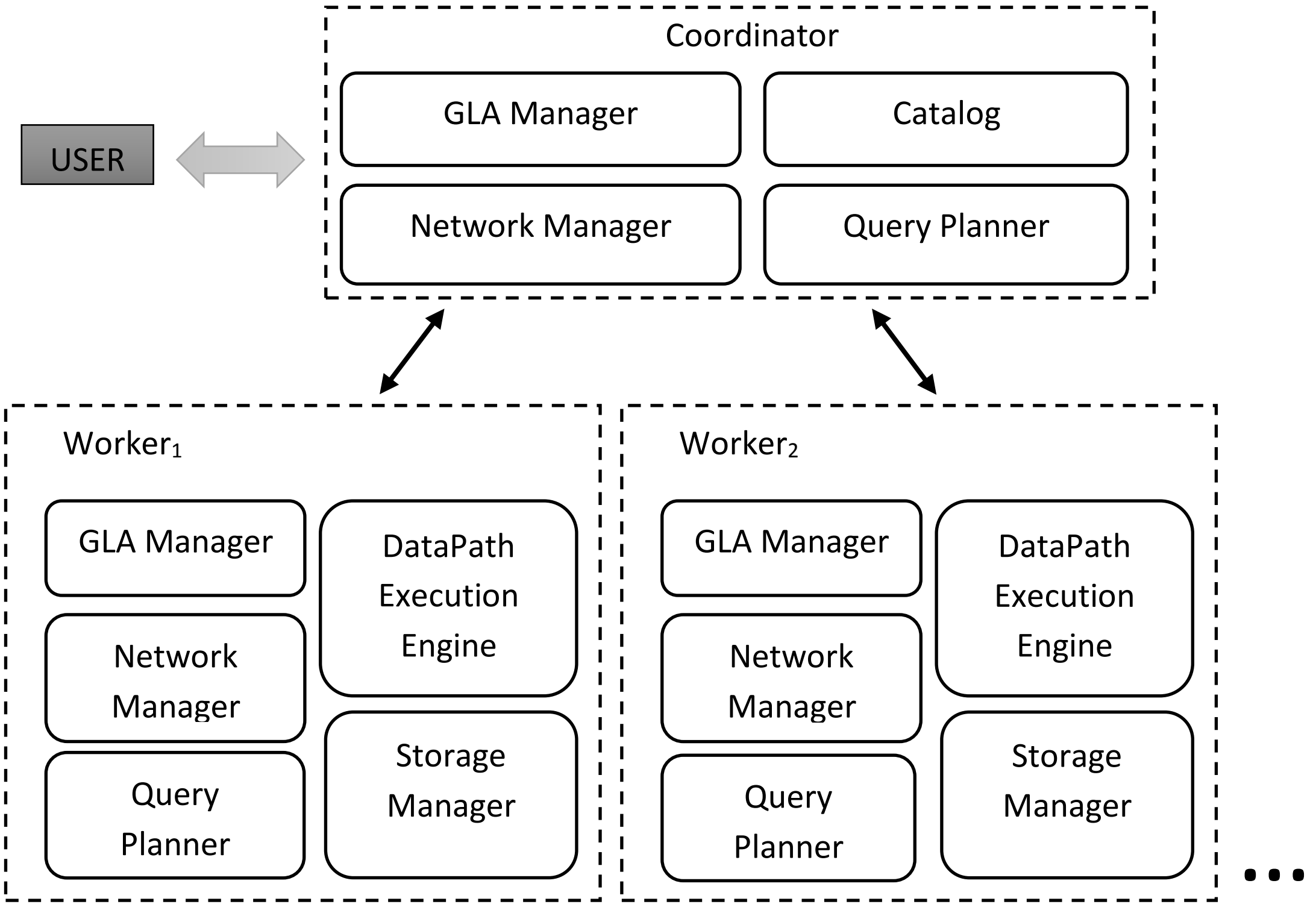

Figure 4.1: GLADE architecture

The way how queries are received and being executed in GLADE is simple. The user is responsible for generation of queries. The queries from user are converted to GLADE plans and then to execution plans which are then executed by the execution engine. The engine consists of 2 different types of nodes on which the execution plan runs, a coordinator node and a worker node. A coordinator works as a scheduler by managing the execution across the network on the worker nodes. The coordinator is incharge of creating a network of workers and replicating the information, like the query execution plan and the network structure, to the worker nodes. Each worker is running aggregates (GLA's) to fetch and direct the data within a node and across the nodes as well. The way how GLA works is explained in the next section. In short, workers extract the data according to the execution plan and process it back to the coordinator, who pases it further back to the user who initiated the query.

GLADE has currently nine clusters, one master node and other eight as a binary tree. Every node can run multiple threads to run GLA's. GLA is executed on a thread if its available and once a GLA has finished its job, it is ready to join a pool of free GLA's.

## 4.2 GLA

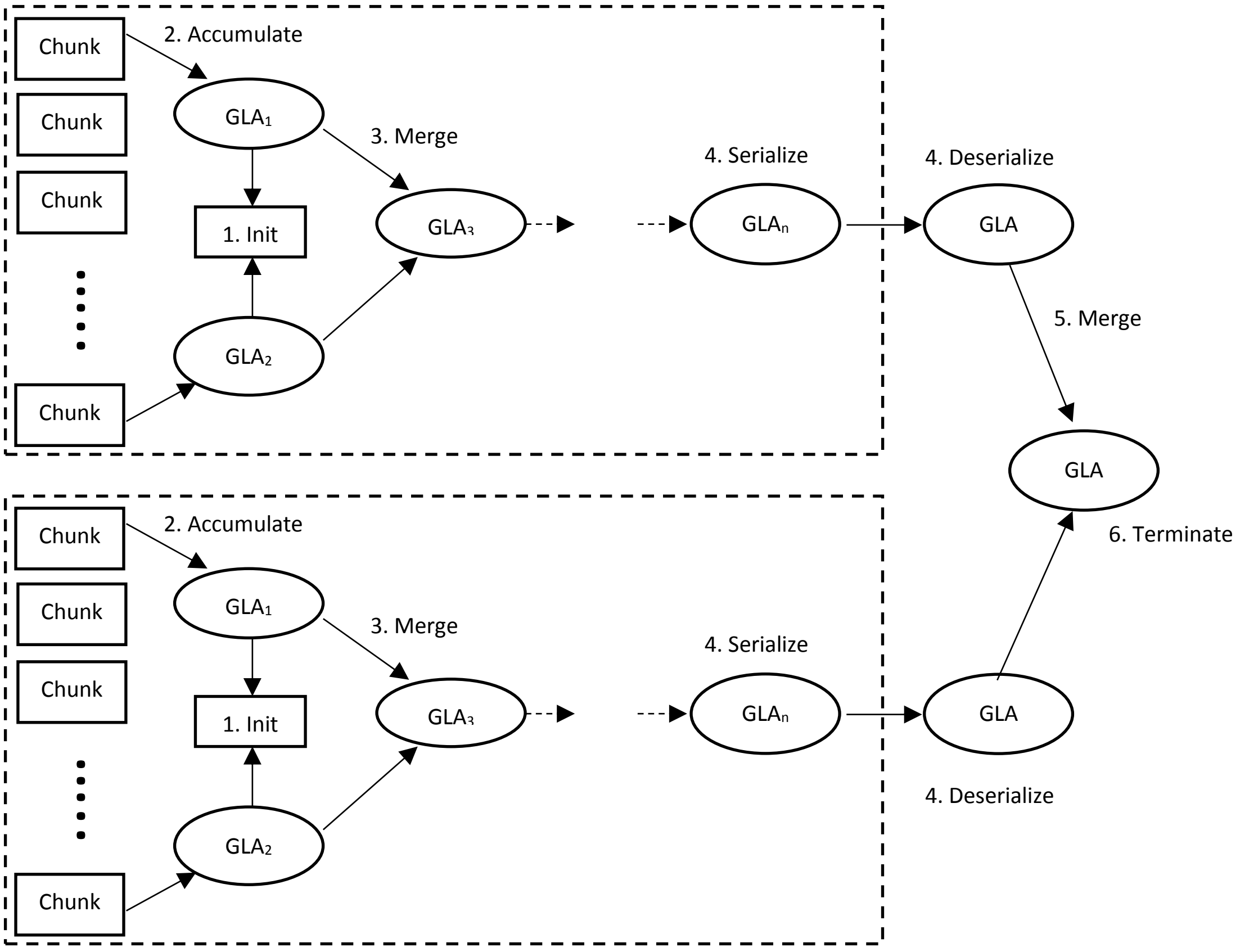

Figure 4.2: GLA Architecture

The architecture of GLA is shown above. Like any other RDBMS, GLADE provides an interface for users to define their own simple functions, or aggregate functions. User defined aggregate functions are a powerful tool to simplify your work by creating functions that are generally not provides by the system by default and which accepts a group of values to generate a sinlge output. They greatly reduce the computation time. Just like in any other RDBMS, UDA in GLADE are typically implemented in the form of a separate class with standard 4 functions to store the **state of the aggregate**. These are initialize, update, merge and finalize functions. In GLADE, we use the following terminology for these functions and their functionality is also explained.

**Init:** The initialize function is used to initialize the state of the UDA. Memory is allocated in initialize state. As the state name shows, all the counters or variables are also initialized here.

**Accumulate:** Accumulate state or the update state keeps updating the state of any UDA. It reads a single tuple at a time and update the state of the aggregate according to the functionality of UDA defined by the user.

**Merge:** Merge, as the name suggests, merges the tuples coming from different nodes. It takes a UDA as input and merges its states with state of current UDA. There can be two types of merging done. One is at the Worker level, which is called Pre-Merging or Pre-Aggregation. This is done in order to minimize the traffic flow over the network from workers to coordinator. The other merging called remote merging at the coordinator level is the final merging which merges all the tuples into a single result.

**Terminate:** This states is called after all the data is being merged and all the computations are being finalized. When a UDA is finished reading tuples, it sends back the updated results to the coordinator for termination. This terminates the UDA. The memory allocated in the Init state is freed in Terminate.

These are the 4 states for any UDA, but GLADE extends this functionality of UDA with two more methods, called Serialize/Deserialize. We call UDA as Generalized Linear Aggregate or GLA in GLADE.

**Serialize/Deserialize:** GLA provides user with the ability to access the state of aggregate directly. Because of the shared-nothing architecture of GLADE, GLAs need to be moved between nodes in order to merge them. This job is done by Serialize/Deserialize method. It is the job of the user to identify in these methods that which variables need to be serialized and how to do that. The way how this works is when a worker finishes the job, it will serialize the GLA and move it to the parent node. At the same time, the deserializer works at the parent node, whose job is to receive and retrieve the GLA. These serialize and deserialize functions needed to implemented in a proper way so that the GLAs can be transferred and retrieved accordingly.

## 4.3 Implementation in GLADE

We have seen in Chapter 3 that how execution tree is generated and how it is transformed into Waypoint and Graph files, which are moved to GLADE for execution of the query inside the query execution engine. The way how these files works is simple. Now we know that how nodes in GLADE works. The execution plan, the Waypoint and Graph files are generated by the Coordinator node. The **Graph file** gives the information that how many queries need to be answered, how many workers needed to be assigned and what jobs each of the worker will be performing. The **Waypoint file** on the other hand helps create the aggregation tree and the complete configuration of the workers network including how the workers are connected and the information of the tuples at each node mainly what tuples are needed and what should be dropped.

After the workers are configured according to the Waypoint and Graph files, each node starts collecting tuples from its local memory and perform computation on it according to the details in the Waypoint file. Once a worker has completed its job, it informs the coordinator. If the worker is the leaf node, it serialized the GLAs for inner nodes. If it is the inner node, it collects the serialized GLAs from other leaves, deserialize it, perform local merge and pass the combined GLA to the parent. The workers at the top finally after receiving all the GLAs from the childs performs one final merge operation and then terminate before sending the final GLA to the coordinator which passes the final results back to the user.

# Chapter 5

# Multi Query Optimization

Last few decades have witnessed an increasing interest in large scale data distribution, data monitoring, and computing. Due to such an exponential rise in the size of data with only limited storage memory, computing environments, processing engines and database management systems have been putting their resources into coming up with renewed areas of distributed query processing and optimization. Perhaps certain issues need to be measured and scrutinized, like answering to a large number of queries over even a bigger sized continuous data streams, SQL-on-Hadoop propaganda, parallel optimization hype and many other. A lot of research is going on right now on query optimization and query processing while considering one query at a time. Query optimizing engines in general tend to work in such a fashion that the processing cost of each query be minimized while it is processed, but individually. We have discusseed in the previous chapters that how optimization plans are generated (in GLADE in our case) and executed for each query separately. But what happens if a set a queries are analyzed simultaneously? In case of single query optimization engines, the overhead evaluated in such manner will be a sum of processing cost of each individual query. This will constitute to an enormous number if the data being processed is of the order of tera bytes or peta bytes, and hundered of redundant queries are being processed with just a slight change in them. To encounter these issues in an optimal way, and to evaluate a single shared plan for all the queries, we present here the Multi Query Optimization (MQO) algorithm by extending our same approach as discussed in Chapter 2.

## 5.1 Introduction

The MQO algorithm is derived from [1] and works on a hybrid of branch and bound and cost based optimization algorithm technique we discussed earlier. Processing a set of queries separately works best if all the queries are discrete and none of the relations are overlapping. However in cases, when a single user is probing various datasets from the same database, chances are there is going to be some overlap. Other cases where processing is done in batches, querying individual queries tends to be inefficient. Although cost of each batch is entirely dependant on the set of queries per batch, the equivalent cost can be greately decreased as compared to running them separately.

MQO algorithm works on generating batches of queries from a pool of queries and then generates a common shared plan for the whole batch. We call a single batch as a Multi Query (MQ) consisting of individual queries where each query is assigned a distinct name. The biggest lead of MQO algorithm on [1] is its capability to incorporate queries into the existing plan at run time as well instead of only generating plans in the offline optimization phase. We will see shortly how this works and what are some of the dependencies of our algorithm. We will explain with examples that how queries are batched and processed in GLADE under MQO algorithm, what are different scenarios that can arise when handeling

Multi Query and how to work with them. We have evaluated our MQ optimizer over TPCH benchmark.

## 5.2 Previous Work

[1] suggests the shared workload optimization algorithm to produce a globally efficient shared access plan for a big workload/set of queries. The heuristic function contains a lot of local minima, dependant on each others. The branch and bound algorithm searches for the solution in the state space with relaxed constraints. The intermediate solutions can be illegal consisting of live or dead nodes or both. The illegal solutions are modified to generate valid solutions.

GLADE is derived from Datapath [2] which is a purely push based database system. This means that queries do not request data, data is instead pushed automatically into the nodes for processing. It deals with analytical queries over multi terabyte data with no indexing or no tunable partitioning. The ultimate goal of [2] is data-centric query processing rather than compute-centric query processing. The flow of data should control the computation rather than the other way around.

Many-Query Join or MQJoin [3] is a novel technique for sharing of join operators while generating plans to effectively deal with hundreds of simultaneousl queries. Their main focus is on minimizing redundant query processing by making efficient use of main memory bandwidth and multi core architechture. While most of the work is done on the impact of sharing on main-memory joins, they used the simple build-probe phased multi-join technique to achieve higher throughput and stable response times.

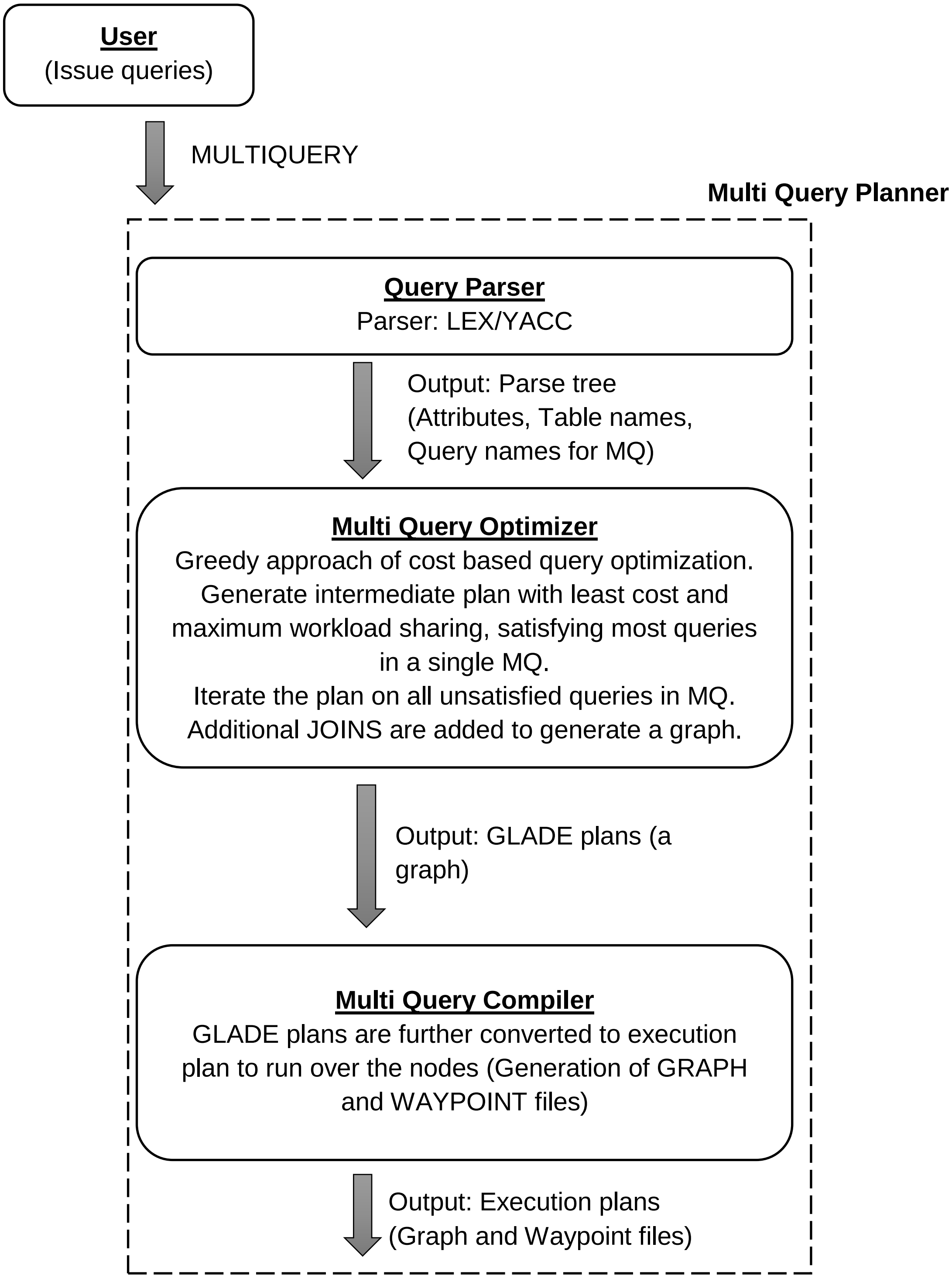

Figure 5.1: Architecture of a Multi Query Planner

## 5.3 Multi Query Planner

MQ planner just like the single query planner discussed in Chapter 2 works on the same technique. Instead of taking a single query, it takes a Multi Query. It also consists of a query parser, optimizer and compiler. The structure of MQ planner is shown in the figure above. In order to understand how MQ planner works, we first need to take a look at the schema of a Multi Query.

### 5.3.1 Multi-Query

```
MULTIQUERY
Query1:
SELECT ..
FROM ..
WHERE ..
Query2:
......
......
......
QueryN:
......
......
END
```

Figure 5.2: Structure of a Multi Query

This is a single MQ. It starts with the keyword “MULTIQUERY”, ends with “END” and inside the block, each query is separately defined starting with a distinct name of the query followed by a colon.

### 5.3.2 Boundry Conditions

Before we explain each part for MQ planner, we need to explain the limitations or bounds of the system. The MQO algorithm works under only one bound, that while we create batches of queries or Multi Queries, the following condition must satisfy:

$$\forall\, \boldsymbol{x}\ (\exists\ \boldsymbol{y} \mid (\boldsymbol{x}, \boldsymbol{y}) \in \boldsymbol{MQ}) \ \Leftrightarrow\ \boldsymbol{T}(\boldsymbol{x}) \ \subseteq\ \boldsymbol{T}(\boldsymbol{y})$$

$$where\ \boldsymbol{T}(\boldsymbol{i}) = Set\ of\ tables\ in\ i^{th} query\ in\ a\ MQ$$

In other words there has to be atleast one query which contains all the distinct tables that are present in all other queries collectively for the same MQ.

## 5.4 Query Parser

In the case of MQ planner, we use the same LEX/YACC query parser but with slight modification. The structure of a single MQ in GLADE has:

- A query name.
- Query data, which is the actual query. Inside query data, we have Select List, Table List and And List for each query.
- A pointer pointing to the next query block.

The way how parser works is that it generates the parse trees in the same way as it generates for a single query, but it is being called *n* number of times, where *n* is the total number of queries in a MQ. Parser generates a linked list of parsed queries along with their names and at the same time, doing 2 other things:

- Semantics Validation in the same way as done earlier.
- Generates a list of all distinct attributes and all distinct tables being used in all the queries collectively.

## 5.5 Multi Query Optimizer

We saw that how query optimizer works for single queries. The cost based optimization approach for generation of a JOINs is very efficient and is definite to generate an optimal join order for all the relational algebra operators with least cost. In case of a single query, we only have one exit point to the user which defines the end of a query. The results are returned to the user via that exit point. However in case of MQ, the scenario is completely different for 2 reasons. First, the query optimizer is not certain if all the queries in a MQ are coming from a single user. Second, there are more than one query to answer regardless of that fact that they are coming from one user or not. So for each query, there has to be a (different) endpoint defined so that in the execution phase, when all tuples are being collected from the database, they can exit via their respective exit points for each query in a MQ. That is why instead of a JOIN tree, we get a JOIN graph with multiple exits, where each exit corresponds to a different query.

## 5.6 Algorithms

Like the cost based optimization algorithm we discussed in Chapter 2, we apply the same strategy on a MQ. A modified dynamic program of CBO with greedy approach is as follows:

**Algorithm 3: Greedy Cost Based MQO Algorithm**

**Input**:

- MQ
- All distinct table names in MQ
- Selection predicates

**Data structures:**

Map : map <table_name> ← tuple (size, cost, order, satisfied queries, valid queries)

- table_name : a name that puts all the table names together, when there are more
- size : the cardinality, i.e., number of tuples, in table(s)
- cost : the sum of intermediate number of tuples in the tree (query optimization cost)
- order : join order with groupings of tables explicit
- satisfied queries: total number of queries satisfied by this join order
- valid queries: list of all the queries satisfied by this join order

**Pre-processing:**

```
for all the individual tables do
        // Initial cost is zero
        if this table satisfies any query in MQ then
                map.insert (table size from catalog, 0, table name, 1, query name)
        end if
        else
                map.insert (table size from catalog, 0, table name, 0, { })
        end else
end for

for all the individual tables do
        for all the queries in MQ do
                // Calculate push down selections for each table among all queries
                // and save it in an array.
        end for
        // Select the largest of all the push down selections from the array for each table
        // and update the map accordingly.
        insert (updated table size after push down selection, …)
        // Don't change other arguments
end for
for all the tables in MQ do
```

**for** all the queries in MQ **do**
// Make a pair of two tables, say $T_1$ and $T_2$, sort ascending $T_1$, $T_2$ (no need
// to have entries for ($T_1$, $T_2$) and ($T_2$, $T_1$) because of commutativity.
// Estimate cardinality of join using distinct values from Catalog and join
// predicates. This is called selectivity.
**end for**

**if** ($T_1$, $T_2$) satisfies any query in MQ **then**
map.insert (Map <($T_1$)>.size * Map <($T_2$)>.size / selectivity, 0, ($T_1$, $T_2$), ++queries satisfied, query name)
**end if**

**else**
map.insert (Map <($T_1$)>.size * Map <($T_2$)>.size / selectivity, 0, ($T_1$, $T_2$), {queries satisfied unchanged}, { })
**end else**

**end for**

// Call the algorithm that generates all possible permutations for all {3, 4, 5 .. n} tables.
GenerateAllPartition(total number of tables, list of tables)

---

This algorithm is different from the CBO algorithm we learned in Chapter 2 in 2 aspects. First we are considering all the queries in MQ. Second we have new added details for our data structures: number of satisfied queries and the list of satisfied queries. The basic idea behind doing this is to optimize the whole process by pruning as much computation as we can. If we have a list of all the satisfied queries for every order, we need not to test the same order again for the same query.

Lets look at how the algorithm works. In the same way as we did in CBO algorithm, we run a loop over all the tables for this MQ and keep a record of the order, size of the table from the catalog, the cost of each table (which is initially zero). For MQ, we test each table for all the queries, since there can be a query which runs over only single table. If there is one for any table, we add 1 to the number of satisfied queries by that particular plan, and write the name of the query to the list of satisfied queries. We do this for all the tables generated by the parser.

The next step is push down selections. Push down selections are different in this case from the original CBO algorithm in the sense that there can be multiple push down selections on the same table in different queries. In order to deal with this clash, we keep a running array of all the queries in which a push down selection is performed on any table. If there are more than one queries, we pick up the highest of them and update the order in the map. The idea behind picking up the highest is that since our ultimate goal is to generate a single join plan for all the queries. To understand this, consider the case when query A generates *n* tuples and query B generates *m* tuples after the selectivity is being performed. Now lets assume that $n > m$. The total number of tuples generated by the two queries all together will be $n + m$ if they were to run indivitually. But since we are generating a single join plan, we will pick *n* number of tuples since we have $n > m$, so all *m* tuples will be present in *n* and if we pick the maximum of $(n, m)$, we can answer both queries with a single selection, hence increasing the overall efficiency.

After performing the selectivity, we continue to follow the same CBO approach and generate a pair for each table. This is slightly different too from actualy CBO algorithm. In case of a MQ, we first pick up a pair of tables and then check that pair of tables across the whole MQ. If there is any query that has a JOIN between these 2 tables, we calculate the selectivity in exact same way as we calculated in CBO for different scenarios we discussed earlier. We keep a list of all the queries that consists of a JOIN of any two tables. At the end of this inner loop, after we calculate all the selectivities for a single pair, we pick the largest selectivity and use that to estimate the size of this JOIN. The same idea is used as discussed above when calculating push down selections. Since we are generating a single join plan, we need to pick $max(\ selectivity\ array\ (T_1, T_2))$. After calculating the selectivity, we need to now check if this JOIN satisfies any query in a MQ (we will do this step every time we generated any new plan at any level). If **this is true**, then we need to add 1 the number of queries satisfied by the each individual plan:

$$Satisfied\ queries\ by\ (T_1,\ T_2) = \begin{cases} Satisfied\ queries\ by\ T_1 + \\ Satisfied\ queries\ by\ T_2 + \\ 1 \end{cases}$$

and do the same thing for the list of satisfied queries for this new JOIN order:

$$List\ of\ queries\ satisfied\ by\ (T_1,\ T_2) = \begin{cases} List\ of\ queries\ satisfied\ by\ T_1 + \\ List\ of\ queries\ satisfied\ by\ T_2 + \\ Query_{(T_1,T_2)} \end{cases}$$

and insert this new entity in the map with the new size for this JOIN.

If **this is false**, we don't have any JOIN of $(T_1,\ T_2)$ in any query in the MQ. The selectivity in that case would be just 1 and the total size of that JOIN will be very expensive and will be equal to the Cartesian product of the cardinalities of individual tables.

After generating all the pairs, the next permutation algorithm is called to create a full state space for all the permutations for all the tables in the MQ. The algorithm is given below:

**Algorithm 4: MQ Partition**

```
Function GenerateAllPartition (N, Array[N])
Input: Total number of tables N, array of tables
    if (Map <sort-asc(T1,T2,...,Tn)> is not empty) then
        // We already have an entry for this combination of tables
        return
    end if
    for i = 1 to N! do  //
         // Get (T1',T2',...,Tn'), the ith permutation of N tables.
        permutation(T1',T2',...,Tn')
        for j = 1 to N-1 do
            left = (T1',...,Tj')
            if (Map <sort-asc(left)> is empty) then
                Partition (j, left)
            end if

            right = (T(j+1)',...,Tn')
            if (Map <sort-asc(right)> is empty) then
               Partition (N-j, right)
            end if

            // At this point we are guaranteed to have entries in the Map on left and right
            // Compute the cost for this tree over all the queries in MQ

            for all the queries in MQ do
                // Check if all the tables on left and right are present in this query in MQ
                // If all tables are present, estimate the cost and size for this JOIN order
                // and store it in an array. We need to pick up the least cost and the
                // maximum size at the end of each iteration.

                if all tables are present then
                    cost = Map <sort-asc(left)>.cost + Map <sort-asc(right)>.cost
                    if (j != 1) then
                        cost += Map <sort-asc(left)>.size
                    end if
                    if (j != N-1) then
                        cost += Map <sort-asc(right)>.size
                    end if
                    if (cost < min_cost) then
```

```
                // estimate cardinality of join using distinct values
                    size = Map <(left)>.size * Map <(right)>.size/selectivity
                    order = Map <sort-asc(left)>.order + Map <sort-asc(right)>.order
                    insert (cost in cost array)
                    insert (size in size array)
                    if this new order satisfies any queries in MQ do
                        // Add 1 to sum of left and right satisfied queries
                        // Add query name to list of satisfied queries
                    end if
                end if
            end if
        end for
        // At this stage we have tested all the queries in MQ for a new JOIN order
        // We need to find the minimum cost now and the maximum size for this JOIN
        min_cost = min(cost array)
        size = min(size array)
    end for     // Go to next permutation
end for     // End of the all permutations for a new JOIN order
Map <sort-asc(T1,T2,...,Tn)> = (size, min_cost, order, satisfied queries, list of satisfied queries)
```

This modified verision of finding all permutations for all the tables works as follows. At any level, we find a new JOIN and find a left and a right node for that JOIN. We then calculate all permutation of JOIN for those tables by generating the permutations using the same lexicographical order algorithm. For any permutation, we have a left node and a right node to join. For both nodes, we run the nodes through all the queries in MQ to find the queries which have a JOIN between the 2 nodes. If any such query exists which contains all the tables in this JOIN, we then calculate the cost and size in the same way as we calculated earlier. Update the number of queries satisfied by this new join order and the number of satisfied queries in the list. We try each permutation on all the queries and after the end of the loop, we pick up the least cost from all the queries since we are interested in finding the least cost. Also we pick the maximum size since we are trying to find a single plan for all the queries. The idea behind picking the max size is the same as we discussed earlier. Once all the permutations are completed, we insert the new JOIN order with least cost and maximum size along with number of satisfied queries into the map.

After this algorithm is completed, the next step is to find out the best possible plan so far from a list of prospective optimal plans. At this stage, we can encounter 2 types of cases which needed to be solved separately. One case will straight generate the final graph file. The other also generates the graph file for the entire MQ but is also responsible for accepting more queries at run time and fitting them to the already existed and running JOIN plan. Before we go into them, lets look at a sample example for each case and explain the further process.

## 5.7 Example Explanation (Case 1)

Consider the given MQ. We will pass the query first through query parser to generate parse trees and to perform semantics validation and to get a list of all total number of tables in it. We can see that the MQ satisfies our limitations and that Query2 includes all the tables involved in all the other queries collectively. The MQ starts with MULTIQUERY keyword and finishes with END keyword and each single query inside the batch follows proper query formatting. For simplicity, we have only considered simple SELECT-FROM-WHERE queries in our test cases. Lets see how this MQ processes through all the steps:

**MULTIQUERY**

*Query1*:

```
SELECT l_orderkey
FROM lineitem
WHERE l_returnflag = 'R' AND l_discount < 0.04 AND l_shipmode = 'MAIL'
```

*Query2*:

```
SELECT l_discount
FROM lineitem, orders, customer, nation, region
WHERE l_orderkey = o_orderkey AND o_custkey = c_custkey AND
        c_nationkey = n_nationkey AND n_regionkey = r_regionkey AND
        r_regionkey = 1 AND o_orderkey < 10000
```

*Query3*:

```
SELECT l_discount
FROM customer, orders, lineitem
WHERE c_custkey = o_custkey AND o_orderkey = l_orderkey AND
        c_name = 'Customer#000070919' AND l_quantity > 30 AND l_discount < 0.03
```

*Query4*:

```
SELECT c_name, c_address, c_acctbal
FROM customer
WHERE c_name = 'Customer#000070919'
```

**END**

The query parser generates all the attributes and all the tables from the MQ as follows:

*List of all the Tables*: lineitem, region, nation, customer, orders

*List of all the Attributes*: l_orderkey, l_returnflag, l_discount, l_shipmode, o_orderkey, o_custkey, c_custkey, c_nationkey, n_nationkey, n_regionkey, r_regionkey, c_name, l_quantity, c_acctbal, c_address

Table encoding generated for this case is:

| | |
|---|---|
| *lineitem* | *0* |
| *region* | *1* |
| *nation* | *2* |
| *customer* | *3* |
| *orders* | *4* |

After scanning the MQ and updating the map, we get:

*0 → (size: 6.00122e+06, cost: 0, queries: 1, valid queries: { Query1 }, order: 0)*

*1 → (size: 5, cost: 0, queries: 0, valid queries: { }, order: 1)*

*2 → (size: 25, cost: 0, queries: 0, valid queries: { }, order: 2)*

*3 → (size: 150000, cost: 0, queries: 1, valid queries: { Query4 }, order: 3)*

*4 → (size: 1.5e+06, cost: 0, queries: 0, valid queries: { }, order: 4)*

After push down selections:

*0 → (size: 6.00122e+06, cost: 0, queries: 1, valid queries: { Query1 }, order: 0)*

*1 → (size: 1, cost: 0, queries: 0, valid queries: { }, order: 1)*

*2 → (size: 25, cost: 0, queries: 0, valid queries: { }, order: 2)*

*3 → (size: 150000, cost: 0, queries: 1, valid queries: { Query4 }, order: 3)*

*4 → (size: 1.5e+06, cost: 0, queries: 0, valid queries: { }, order: 4)*

Generating table pairs:

| **Relation** | **Size** | **Cost** | **Queries** | **Valid Queries** | **Join Order** |
|---:|---:|---:|---:|---:|---:|
| 0 | 6.00122e+06 | 0 | 1 | Query1 | 0 |
| 01 | 6.00122e+06 | 0 | 1 | Query1 | 01 |
| 02 | 1.5003e+08 | 0 | 1 | Query1 | 02 |
| 03 | 9.00182e+11 | 0 | 2 | Query1, Query4 | 03 |
| 04 | 6.00122e+06 | 0 | 1 | Query1 | 04 |
| 1 | 1 | 0 | 0 | { } | 1 |
| 12 | 5 | 0 | 0 | { } | 12 |
| 13 | 150000 | 0 | 1 | Query4 | 13 |
| 14 | 1.5e+06 | 0 | 0 | { } | 14 |
| 2 | 25 | 0 | 0 | { } | 2 |
| 23 | 150000 | 0 | 1 | Query4 | 23 |
| 24 | 3.75e+07 | 0 | 0 | { } | 24 |
| 3 | 150000 | 0 | 1 | Query4 | 3 |
| 34 | 1.5e+06 | 0 | 1 | Query4 | 34 |
| 4 | 1.5e+06 | 0 | 0 | { } | 4 |

When we look at the size, we can see very big numbers which corresponds to the Cartesian product of two relations. The main reason for this is the absence of such JOIN from most of the queries in a MQ.

A full table with all permutations generated by the algorithm is:

| **Relation** | **Size** | **Cost** | **Queries** | **Valid Queries** | **Join Order** |
|---|---|---|---|---|---|
| 0 | 6.00122e+06 | 0 | 1 | Query1 | 0 |
| 01 | 6.00122e+06 | 0 | 1 | Query1 | 01 |
| 012 | 3.00061e+07 | 5 | 1 | Query1 | 0(12) |
| 0123 | 1.80036e+11 | 30005 | 2 | Query1, Query4 | 0((12)3) |
| **01234** | **1.20024e+06** | **330005** | **3** | **Query1, Query2, Query4** | **0(((12)3)4)** |
| 0124 | 3.00061e+07 | 6.00122e+06 | 1 | Query1 | (12)(04) |
| 013 | 9.00182e+11 | 150000 | 2 | Query1, Query4 | 0(13) |
| 0134 | 6.00122e+06 | 1.65e+06 | 2 | Query1, Query4 | 0((13)4) |
| 014 | 6.00122e+06 | 1.5e+06 | 1 | Query1 | 0(14) |
| 02 | 1.5003e+08 | 0 | 1 | Query1 | 02 |
| 023 | 9.00182e+11 | 150000 | 2 | Query1, Query4 | (23)0 |
| 0234 | 6.00122e+06 | 1.65e+06 | 2 | Query1, Query4 | 0(4(23)) |
| 024 | 1.5003e+08 | 6.00122e+06 | 1 | Query1 | 2(04) |
| 03 | 9.00182e+11 | 0 | 2 | Query1, Query4 | 03 |
| **034** | **6.00122e+06** | **1.5e+06** | **3** | **Query1, Query3, Query4** | **0(34)** |
| 04 | 6.00122e+06 | 0 | 1 | Query1 | 04 |
| 1 | 1 | 0 | 0 | { } | 1 |
| 12 | 5 | 0 | 0 | { } | 12 |
| 123 | 30000 | 5 | 1 | Query4 | (12)3 |
| 1234 | 300000 | 30005 | 1 | Query4 | ((12)3)4 |
| 124 | 7.5e+06 | 5 | 0 | { } | 4(12) |
| 13 | 150000 | 0 | 1 | Query4 | 13 |
| 134 | 1.5e+06 | 150000 | 1 | Query4 | (13)4 |
| 14 | 1.5e+06 | 0 | 0 | { } | 14 |
| 2 | 25 | 0 | 0 | { } | 2 |
| 23 | 150000 | 0 | 1 | Query4 | 23 |
| 234 | 1.5e+06 | 150000 | 1 | Query4 | 4(23) |
| 24 | 3.75e+07 | 0 | 0 | { } | 24 |
| 3 | 150000 | 0 | 1 | Query4 | 3 |
| 34 | 1.5e+06 | 0 | 1 | Query4 | 34 |
| 4 | 1.5e+06 | 0 | 0 | { } | 4 |

At this stage, when all the permutations are generated, the next step is to pick up the best possible join order. The criteria to pick up the best plan is two fold. First we pick up the best plan from all the plan that **satisfies the most number of queries**. Second, if there more than such cases, we pick the one with the **least cost**. Since our ultimate goal now is to find a single plan that can answer all the queries, the cost becomes the second parameter in deciding which join order to choose.

In this above example, we have the case where we end up getting 2 choices which satisfies 3 queries: join order *0(34)* and *0(((12)3)4).* After sampling these two rules from the set of all other rules, we finally pick up the one with the least cost, which is *0(((12)3)4)*. This join order at this stage is definite to have the least cost and satisfies maximum queries in a MQ and is the most optimal join order to start our next algorithm for generation of the entire graph. We now discuss the next algorithm which generates the final shared optimized plan for the MQ.

**Algorithm 5: Shared Optimized Plan Genration**

```
Function sharedOptimizedPlan ( MQ _mq,  OptimalJoinPlan p)
if p.validQueries = total number of queries in MQ then
        // All queries are already satisfied by this plan. No need to test next query in MQ.
        return
end if
Query q = randomWalk (_mq)
        // For the current unsatisfied query, we need to check 2 things:
        // 1. Either the best plan we found has all the tables in it, in that case
        //    we need to add more operators only by duplicating the tables (Case 1)
        // 2. Either the best plan we found has less than total number of tables,
        //    in that case we need to add more joins on top of that by joining that
        //    with more tables which are missing (Case 2 in next section)
switch (q)
        case 1:
                addMoreJoins (q, p)
        case 2:
                addMoreTables(q, p)
end switch
sharedOptimizedPlan ( _mq,  p)
```

We can see that Algorithm 4 picked 2 possible options which satisfy maximum number of queries. We picked the one with least cost, and the join order contains all the distinct tables in it. It could be the case (we will discuss in the next example) that the algorithm picks up the join order which will have less tables in it. We will see that case in the next section and see what are the benefits of that algorithm. We will consider the first algorithm here.

The shared optimized plan generation algorithm performs a **random walk** on the unsatisfied queries and picks a random numbered unsatisfied query from the pool of unsatisfied queries and apply the cases on that selected query. Since in the case of this MQ, we only are left with one unsatisfied query, so the algorithm will pick that query.

**Algorithm 5: Duplicating JOINS in Optimization Tree**

**Function addMoreJoins ( Query q,  OptimalJoinPlan p)**
// First we need to generate an optimization for this single query alone
CBO(q)
// If this is the first time in this loop, the graph is not generated yet for the initial
// join order, so we generate an initial graph for the optimized join order just like
// the same way we generated the optimization tree in case of CBO
**if** graph is not created **then**
    graphGenerator(p.JoinOrder)
**end if**
// Now we need to recursively check if the small tree (or its sub-trees) are a sub-tree of
// the large graph file or not

recursiveSubTreeChecking (root of graph, root of Query q)
// After return from this function, the query q is added in the graph, we need to
// keep a list of all the exits for each query

exits.insert(root of q)
**return**

**Function recursiveSubTreeChecking (Tree $T_1$, Tree $T_2$)**
**if** $T_2$ is null **then**
    // A null is always a sub tree of any tree
    **return true**
**end if**
boolean left ← false, right ← false
left = subTree ($T_1$, $T_2$.leftChild)
**if** left is false **then**
    recursiveSubTreeChecking($T_1$, $T_2$.leftChild)
**end if**
right = subtree ($T_1$, $T_2$.rightChild)
**if** right is false **then**
    recursiveSubTreeChecking($T_1$, $T_2$.rightChild)
**end if**
**if** both left and right are true **then**
    // That means both left and right nodes are a sub tree of big graph.
    // Find the place in the graph and add both left and right nodes on top of that.
    findToAdd($T_1$, $T_2$.leftChild)
    findToAdd($T_1$, $T_2$.rightChild)

```
end if
if left is true but right is false then
        // Only left child is a node of graph. Find the place to add in the graph.
        findToAdd(T1, T2.leftChild)
end if
if left is false but right is true then
        // Only right child is a node of graph. Find the place to add in the graph.
        findToAdd(T1, T2.rightChild)
end if

Function subTree (Tree T1, Tree T2)
if T2 is null then
        // The empty tree is always a sub tree
        return true
end if
if T1 is null then
        // Big tree is empty and still sub tree is not found, so no subtree exists
        return false
end if
if T1.tables = T2.tables and if T1.order = T2.order then
        return true
end if
return (subTree (T1.leftChild, T2) or subTree (T1.rightChild, T2))
```

The algorithm starts with generating an optimization tree for a single query. Since this query is not satisfied by the optimal join order plan, the optimal join plan for this query will most certainly not satisfy the join plan either. The next step in graph generation is the graph initiation. If this is the first time the program has entered this loop, then we have not started generating the graph yet. So the initial optimization tree is generated for the optimal join order in the exact same way as we generated in Chapter 2. For the case of our example MQ, we have the optimal join order as *0(((12)3)4)* which satisfies 3 queries, Query1, 2 and 4. The only query it does not satisfy is Query3. We have Query3 as:

**SELECT** l_discount

**FROM** customer, orders, lineitem

**WHERE** c_custkey = o_custkey **AND** o_orderkey = l_orderkey **AND**

c_name = 'Customer#000070919' **AND** l_quantity > 30 **AND** l_discount < 0.03

We pass this query to the CBO and it generated the optimization tree for this query. The optimal join order for this query comes out to be *0(34)*, which is shown by the following optimization tree:

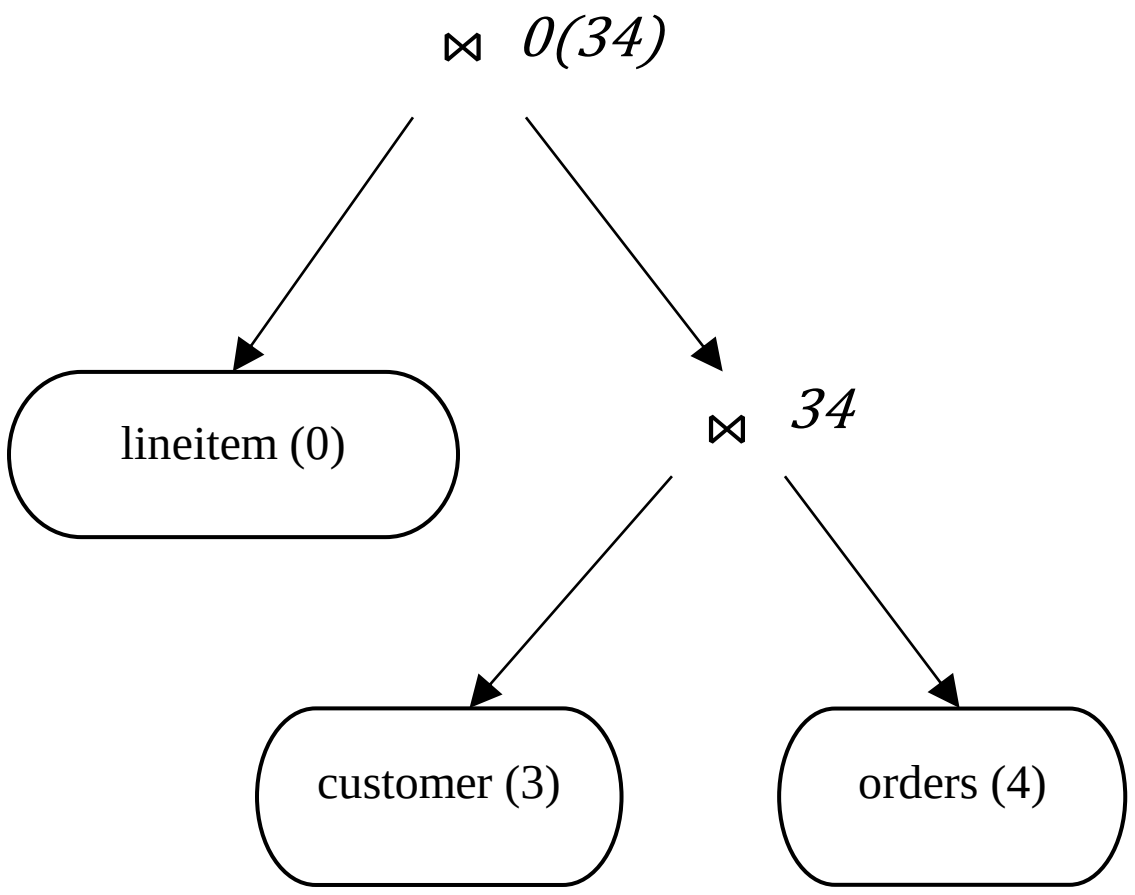

Figure 5.3: JOIN order for unsatisfied in MQO query generated by CBO

The next step is to generate the initial graph for the optimal join order *0(((12)3)4)*. This graph generation is exactly the same as tree generation in the case of CBO. The only difference is that during graph generation, while we are generating the graph at each level, we keep on cheking if each level satisfies a query in MQ or not, since we want to keep a track of all the exit points for the MQ. Whenever we encounter an exit point, we log that exit point in a global list of exits.

The graph generated for *0(((12)3)4)* is as follows:

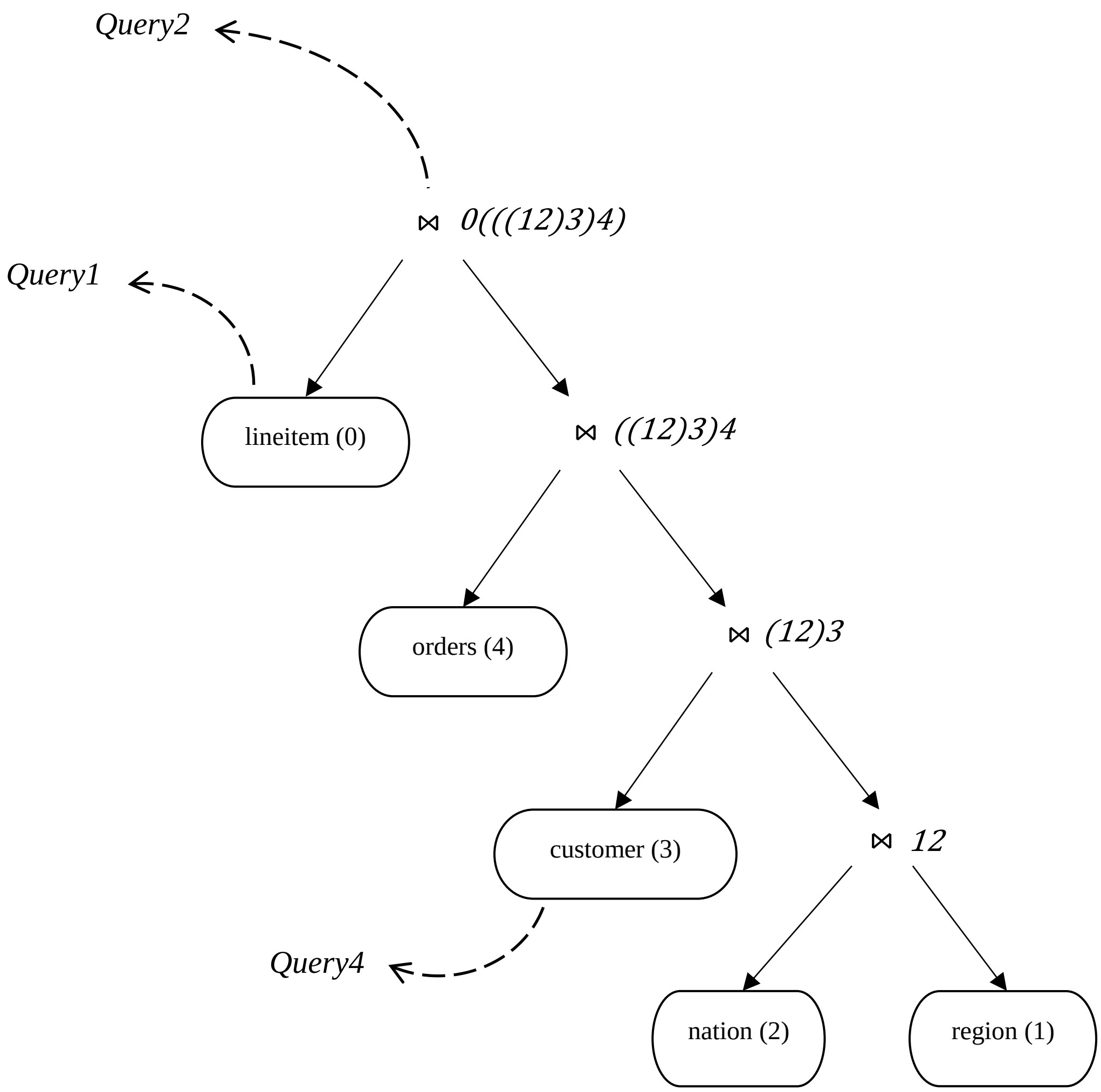

Figure 5.4: Graph with multiple exit points, each corresponding to a single query in a MQ

Once the graph is generated, the next step is to add all the unsatisfied queries to the graph one by one. We can see in the graph that all the dotted lines show the exit points, from where each query exits in a MQ. We also keep in mind all the rules for generating a graph, that the small node always goes to the right side, and each node contains the informarion about all the subnodes from that level, number of parents and information about parent nodes and if any particular node is an exit point or not. We pass this graph and the above tree for the unsatisfied query to the recursive sub tree checking routine, which checks if the small tree or any sub tree of that is another sub tree of large graph file, and if it is, find the spot where to affix it.

Recursive sub tree checking routine recursively splits the tree and check in the graph file if it is a sub tree of the graph file. In order to get a better idea about this, consider the example MQ. We have the tree as *0(34)* and the graph as *0(((12)3)4)*. This recursive sub tree checking routine splits *0(34)* into two parts: left *(0)* and right *(34)*. We go to the right first and see that *34* is not a sub tree of the graph *0(((12)3)4)*. So the routine further splits *34* into left *(3)* and right *(4)*. Now both left and right are sub trees of the graph. At this stage, the algorithm traverses the graph to find the place where node *(3)* is present in the graph. It adds another parent pointer to that node *(3)* and creates a new JOIN node on top of that and assigns this node *(3)* as the left child of the new JOIN node. The algorithm then traverses for the other node *(4)* and follows the process and assigns node 4 as the right child of the newly created JOIN node. It then returns the parent pointer of the new JOIN node back to the previous call of the function. This scenario of adding a new JOIN node is shown below:

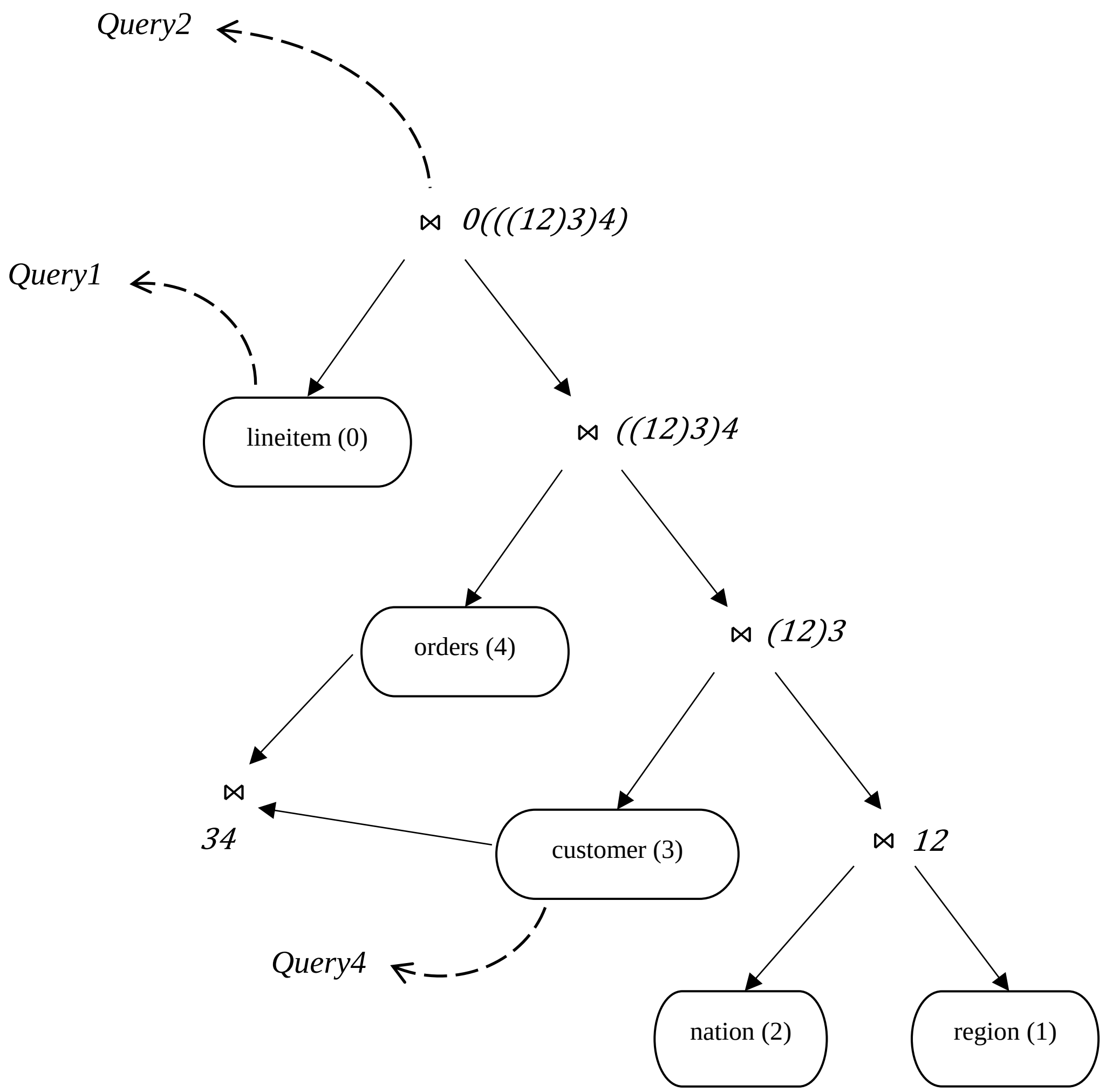

Figure 5.5: Intermediate step in generation of complete graph for a MQ

The algorithm goes back to the previous call, where it finds that now, the node *(0)* from the tree is also a sub tree of the graph. So it traverses the graph to find node *(0)* and creates a new JOIN node on top of that, assigns node *(0)* as the left child of this new node and assigns the previously returned JOIN node as the right child of this new JOIN node. This is shown below:

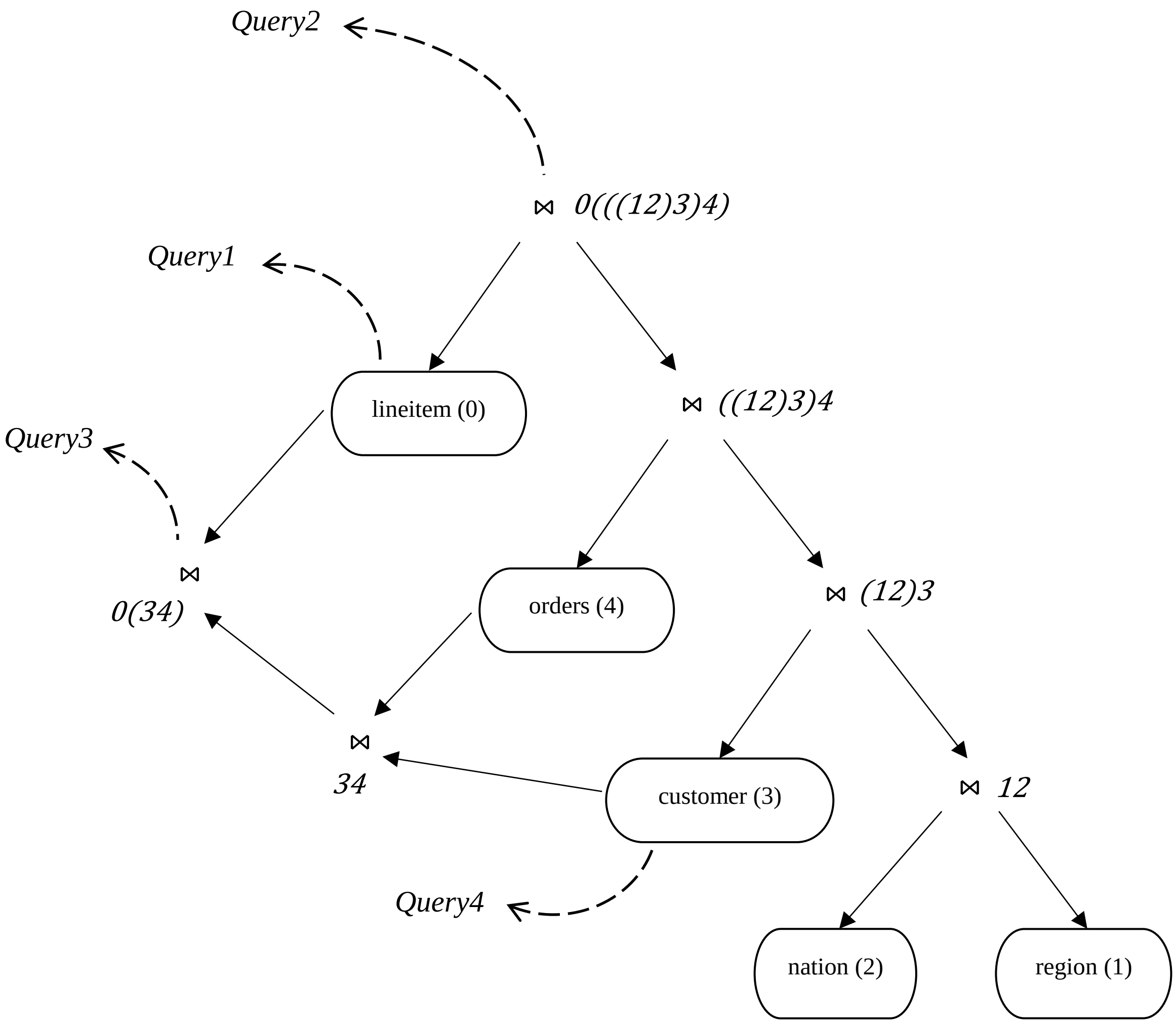

Figure 5.6: A complete graph for a MQ satisfying all queries generated by MQO when added more JOINS

At this stage, the recursive sub tree checking routine is done with adjusting the tree in the graph, so the routine returns back to “add more joins” routine. It then updates the number of queries satisfied by this new plan, updates the exits list by adding a newly created exit for this query and terminate the case. The algorithm on top of that (shared optimized plan generation) loops back again and finds out that now, total number of queries satisfied by the current plan is equal to the total number of queries. Hence the MQO algorithm terminates with the graph file shown above as the final single optimal join plan to satisfy all the queries in the MQ. This graph file is then passed on to the query compiler for further generation of graph and waypoint files needed to be executed by GLADE to start generating tuples for this MQ.

## 5.8 Example Explanation (Case 2)

We discussed in Shared optimized plan generation algorithm that there can be 2 cases which can arise when deciding for the best possible join order we get from Generate All Partition routine. We have already discussed Case 1 when the best join plan has all the tables included. Now consider this MQ which will generate a different case scenario. Again in the same way, we will pass the query first through query parser to generate parse trees and to perform semantics validation and to get a list of all total number of tables in it. We will go through all the steps exactly in the same way as above to find out what we get in the end and how it is different from Case 1 and how to tackle Case 2.

**MULTIQUERY**

*Query1:*

**SELECT** l_orderkey

**FROM** lineitem

**WHERE** l_returnflag = 'R' **AND** l_discount < 0.04 **AND** l_shipmode = 'MAIL'

*Query2:*

**SELECT** l_discount

**FROM** lineitem, orders, customer, nation, region

**WHERE** l_orderkey = o_orderkey **AND** o_custkey = c_custkey **AND** c_nationkey = n_nationkey

**AND** n_regionkey = r_regionkey **AND** r_regionkey = 1 **AND** o_orderkey < 10000

*Query3:*

**SELECT** l_discount

**FROM** customer, orders, lineitem

**WHERE** c_custkey = o_custkey **AND** o_orderkey = l_orderkey **AND**

c_name = 'Customer#000070919' **AND** l_quantity > 30 **AND** l_discount < 0.03

*Query4:*

**SELECT** c_name, c_address, c_acctbal

**FROM** customer

**WHERE** c_name = 'Customer#000070919'

*Query5:*

**SELECT** c_name

**FROM** customer, orders

**WHERE** c_custkey = o_custkey **AND** o_totalprice < 10000

**END**

The query parser generates all the attributes and all the tables from the MQ as follows:

*List of all the Tables*: lineitem, region, nation, customer, orders

*List of all the Attributes*: l_orderkey, l_returnflag, l_discount, l_shipmode, o_orderkey, o_custkey, c_custkey, c_nationkey, n_nationkey, n_regionkey, r_regionkey, c_name, l_quantity, c_acctbal, c_address, o_totalprice

Table encoding generated for this case is:

| | |
|---|---|
| *lineitem* | *0* |
| *region* | *1* |
| *nation* | *2* |
| *customer* | *3* |
| *orders* | *4* |

After scanning the MQ and updating the map, we get:

*0 → (size: 6.00122e+06, cost: 0, queries: 1, valid queries: { Query1 }, order: 0)*

*1 → (size: 5, cost: 0, queries: 0, valid queries: { }, order: 1)*

*2 → (size: 25, cost: 0, queries: 0, valid queries: { }, order: 2)*

*3 → (size: 150000, cost: 0, queries: 1, valid queries: { Query4 }, order: 3)*

*4 → (size: 1.5e+06, cost: 0, queries: 0, valid queries: { }, order: 4)*

After push down selections:

*0 → (size: 6.00122e+06, cost: 0, queries: 1, valid queries: { Query1 }, order: 0)*

*1 → (size: 1, cost: 0, queries: 0, valid queries: { }, order: 1)*

*2 → (size: 25, cost: 0, queries: 0, valid queries: { }, order: 2)*

*3 → (size: 150000, cost: 0, queries: 1, valid queries: { Query4 }, order: 3)*

*4 → (size: 1.5e+06, cost: 0, queries: 0, valid queries: { }, order: 4)*

Generating table pairs and a full table with all permutations generated by the algorithm:

| **Relation** | **Size** | **Cost** | **Queries** | **Valid Queries** | **Join Order** |
|---|---|---|---|---|---|
| 0 | 6.00122e+06 | 0 | 1 | Query1 | 0 |
| 01 | 6.00122e+06 | 0 | 1 | Query1 | 01 |
| 012 | 3.00061e+07 | 5 | 1 | Query1 | 0(12) |
| 0123 | 1.80036e+11 | 30005 | 2 | Query1, Query4 | 0((12)3) |
| 01234 | 1.20024e+06 | 330005 | 3 | Query1, Query2, Query4 | 0(((12)3)4) |
| 0124 | 3.00061e+07 | 6.00122e+06 | 1 | Query1 | (12)(04) |
| 013 | 9.00182e+11 | 150000 | 2 | Query1, Query4 | 0(13) |
| 0134 | 6.00122e+06 | 1.65e+06 | 2 | Query1, Query4 | 0((13)4) |
| 014 | 6.00122e+06 | 1.5e+06 | 1 | Query1 | 0(14) |
| 02 | 1.5003e+08 | 0 | 1 | Query1 | 02 |
| 023 | 9.00182e+11 | 150000 | 2 | Query1, Query4 | (23)0 |
| 0234 | 6.00122e+06 | 1.65e+06 | 2 | Query1, Query4 | 0(4(23)) |
| 024 | 1.5003e+08 | 6.00122e+06 | 1 | Query1 | 2(04) |
| 03 | 9.00182e+11 | 0 | 2 | Query1, Query4 | 03 |
| 034 | 6.00122e+06 | 1.5e+06 | 4 | Query1, Query3, Query4, Query5 | 0(34) |
| 04 | 6.00122e+06 | 0 | 1 | Query1 | 04 |
| 1 | 1 | 0 | 0 | { } | 1 |
| 12 | 5 | 0 | 0 | { } | 12 |
| 123 | 30000 | 5 | 1 | Query4 | (12)3 |
| 1234 | 300000 | 30005 | 1 | Query4 | ((12)3)4 |
| 124 | 7.5e+06 | 5 | 0 | { } | 4(12) |
| 13 | 150000 | 0 | 1 | Query4 | 13 |
| 134 | 1.5e+06 | 150000 | 1 | Query4 | (13)4 |
| 14 | 1.5e+06 | 0 | 0 | { } | 14 |
| 2 | 25 | 0 | 0 | { } | 2 |
| 23 | 150000 | 0 | 1 | Query4 | 23 |
| 234 | 1.5e+06 | 150000 | 1 | Query4 | 4(23) |
| 24 | 3.75e+07 | 0 | 0 | { } | 24 |
| 3 | 150000 | 0 | 1 | Query4 | 3 |
| 34 | 1.5e+06 | 0 | 2 | Query4, Query5 | 34 |
| 4 | 1.5e+06 | 0 | 0 | { } | 4 |

Now in this case, when selecting best plan, the algorithm picks *0(34)* join order which satisfies 4 queries. This join order contains three tables only, whereas total number of tables are five. We need to add more tables on top of this join order for all the unsatisfied queries. The algorithm for that is “add more tables” routine, which is defined here:

**Algorithm 6: Adding Missing Tables in Optimization Tree**

**Function addMoreTables ( Query q,  OptimalJoinPlan p)**
**if** graph is not created **then**
    graphGenerator(p.JoinOrder)
**end if**
// Find the list of tables missing in the join order from this query.
// For each table in list, insert in into a new map along with all the details
// and treat the join order as a single relation, say *r* and insert it into the map too.
// Then run the CBO algorithm on this new map and replace the final best plan with p.
**for** all tables in q **do**
    **if** table is not in p **then**
        new_map.insert (table size from catalog, 0, table name, 0, { })
    **end if**
**end for**
new_map.insert (p)
CBO (new_map)
new_p = new_map.find (best join plan)
old_map.replace (p with new_p)
exits.insert (q)
**return**

The first step in this algorithm is to generate a graph from the join order if it has not been generated yet. This will create all the exit points for all the satisfying queries at the current stage. After that, the algorithm finds all the tables that are missing from this join order, creates a new map and inserts all of those tables in the map with initial cost and sizes and any satisfied queries, if exists. The next step is to transform the current join order into a single relation or treat it as a view, where all the details are hidden underneath it. Insert that view into the new map too. Now we have a single map with single enteties and we can run our CBO algorithm on it, treating it as a simple query. The only difference will be that the starting cost of this join order in the form of a view will be the actual cost of the join order generated by MQ optimizer.

Let us follow the example to have a better picture of the algorithm. The optimal join order is *0(34)* which satisfies Query 1, 3, 4 and 5. The only unsatisfied query is Query2. Running this algorithm starting with Query2 and the join order *0(34)*, the algorithm creates a list of missing tables, which are *(1,2)*. We treat *0(34)* as a virtual view, we call it *v* with join order as *0(34)* and satisfying 4 queries. The starting map will look like:

*1 → size: 1, cost: 0, queries: 0, valid queries: { }, order: 1*

*2 → size: 25, cost: 0, queries: 0, valid queries: { }, order: 2*

*v → size: 6.00122e+06, cost: 1.5e+06, queries: 4, valid queries: { Query1, Query3, Query4, Query5 }, order: (0(34))*

Now we will run CBO algorithm on this. After the push down selections and the table pairings, we will get:

| **Relation** | **Size** | **Cost** | **Queries** | **Valid Queries** | **Join Order** |
|---:|---:|---:|---:|---:|---:|
| 1 | 1 | 0 | 0 | { } | 1 |
| 12 | 25 | 0 | 0 | { } | 12 |
| 1v | 6.00122e+06 | 1.5e+06 | 4 | Query1, Query3, Query4, Query5 | 1(0(34)) |
| 2 | 25 | 0 | 0 | { } | 2 |
| 2v | 6.00122e+06 | 1.5e+06 | 4 | Query1, Query3, Query4, Query5 | 2(0(34)) |
| v | 6.00122e+06 | 1.5e+06 | 4 | Query1, Query3, Query4, Query5 | (0(34)) |

And after running the complete algorithm, the full table is generated, which is given below:

| **Relation** | **Size** | **Cost** | **Queries** | **Valid Queries** | **Join Order** |
|---|---|---|---|---|---|
| 1 | 1 | 0 | 0 | { } | 1 |
| 12 | 25 | 0 | 0 | { } | 12 |
| 12v | 1.20024e+06 | 1.5e+06 | 5 | Query1, Query2, Query3, Query4, Query5 | (12)(0(34)) |
| 1v | 6.00122e+06 | 1.5e+06 | 4 | Query1, Query3, Query4, Query5 | 1(0(34)) |
| 2 | 25 | 0 | 0 | { } | 2 |
| 2v | 6.00122e+06 | 1.5e+06 | 4 | Query1, Query3, Query4, Query5 | 2(0(34)) |
| v | 6.00122e+06 | 1.5e+06 | 4 | Query1, Query3, Query4, Query5 | (0(34)) |

This table contains the entity <missing tables + *v*> or in the case of this example, it is *12v,* which will satisfy this new query after adding the missing tables in this query on top of the actual join order, which was $v = 0(34)$. The next step is to find that what is the relation index of this new entity. In this case it is *12v.* Replace that entity in the actual map with these new result. The old entity in the acutal table was:

*01234 → size: 1.20024e+06, cost: 330005, queries: 3, valid queries: { Query1, Query2, Query4 }, order: 0(((12)3)4)*

which is replaced by:

*01234 → size: 1.20024e+06, cost: 1.5e+06, queries: 5, valid queries: { Query1, Query5, Query3, Query2, Query4 }, order: (12)(0(34))*

The last step is to insert this exit in the list of exits as we have found another satisfying query in the graph which corresponds to another exit point in the grap. This now terminates our algorithm and the program loops back to Shared Optimized Plan Genration algorithm and then finds out that now total number of queries are equal to the plan p, hence it terminates the algorithm with final join order of the graph as *(12)(0(34))* with 5 exits points, each corresponding to a single query. The graph representation of the whole process is shown below:

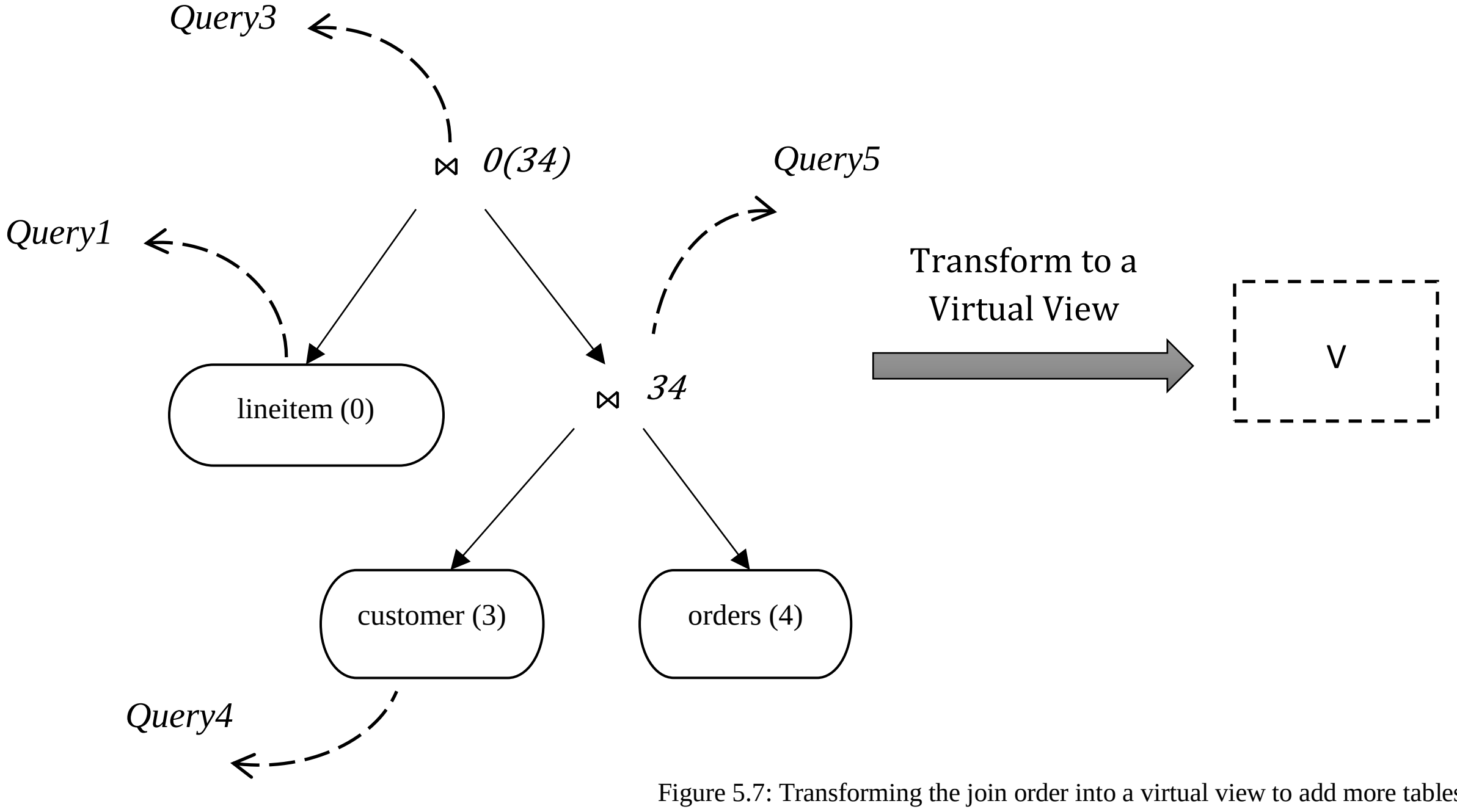

Figure 5.7: Transforming the join order into a virtual view to add more tables

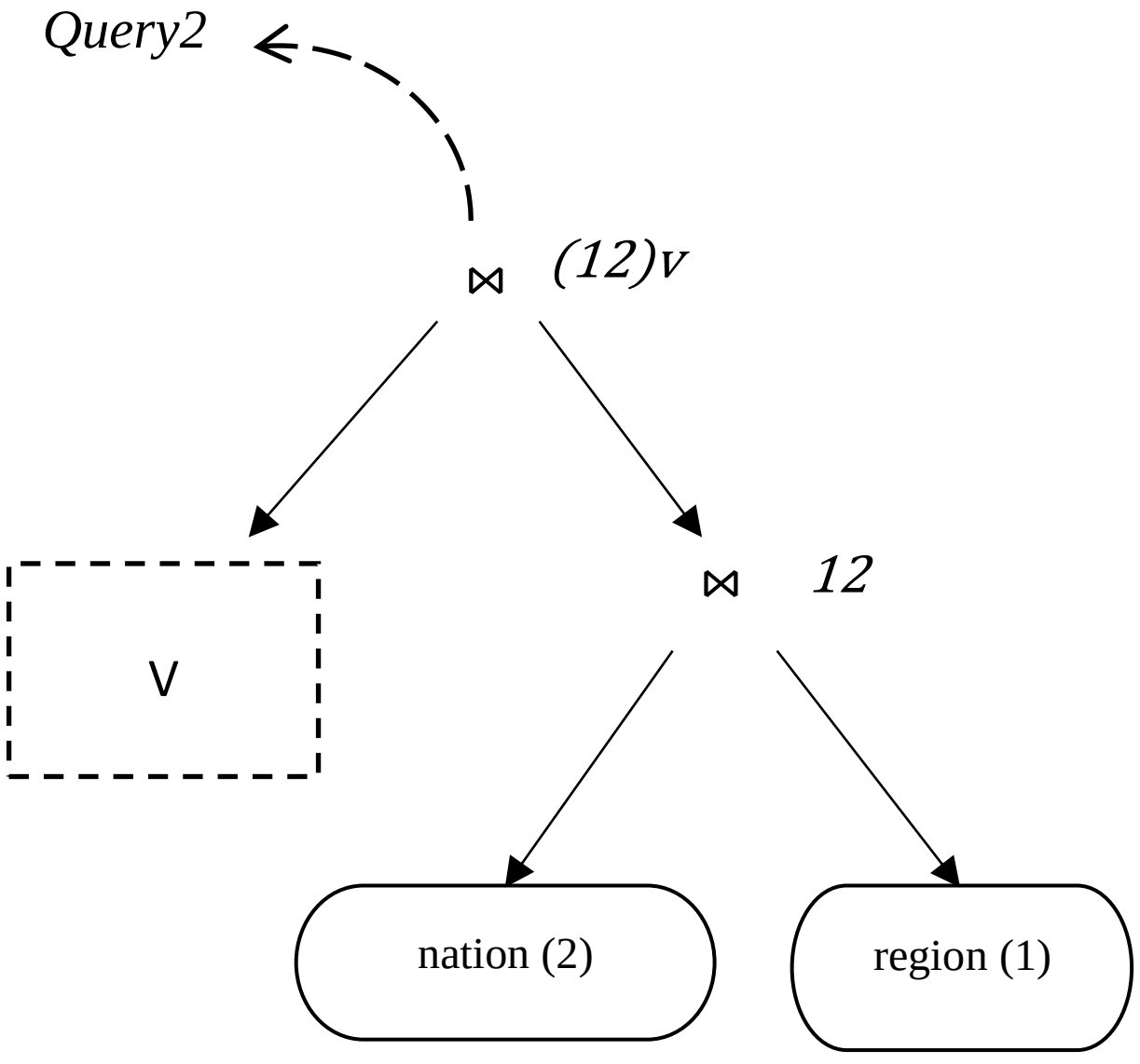

Figure 5.8: Simple tree generated by CBO which contains a a graph represented as a view "V"

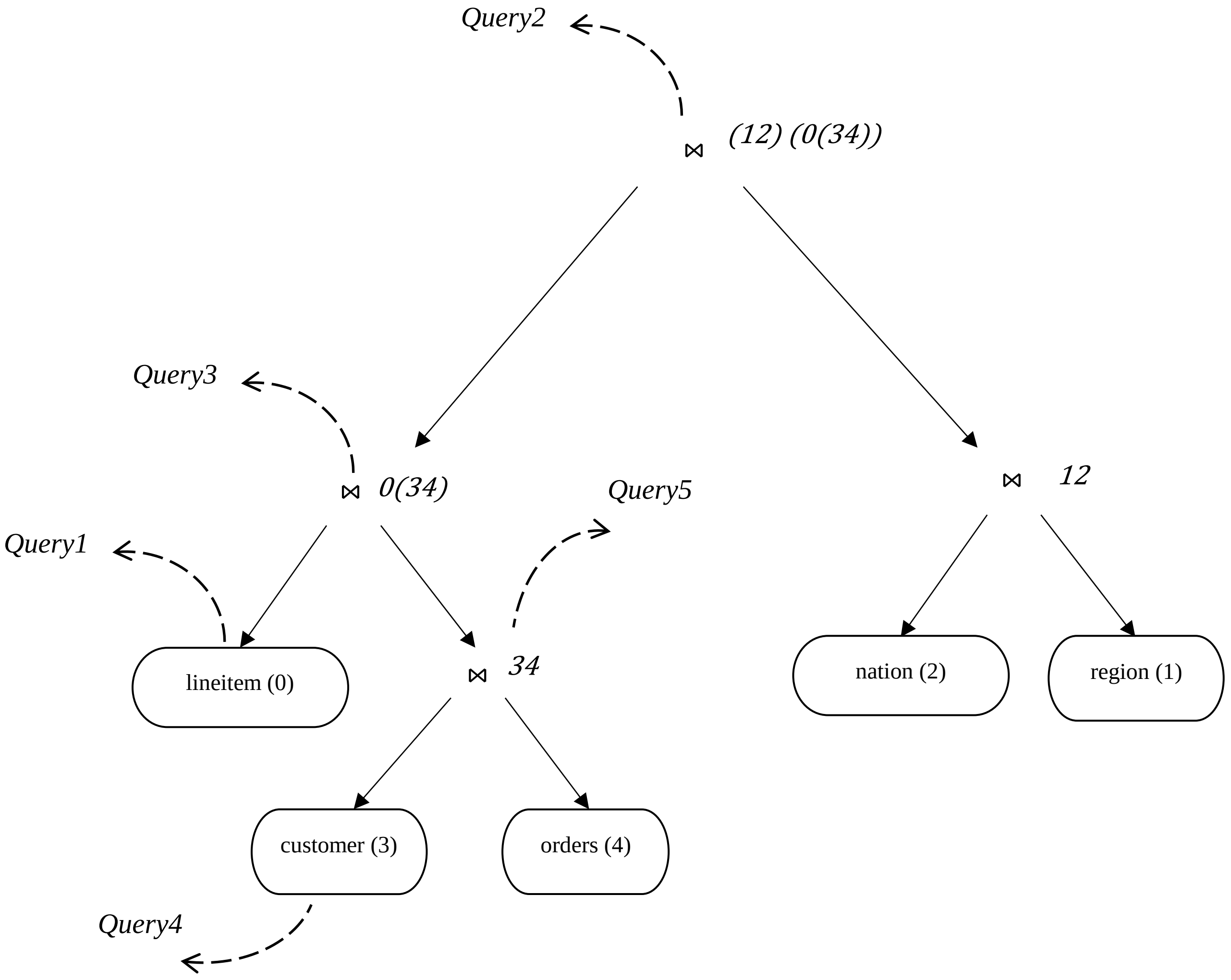

Figure 5.9: Final graph generated by MQO for Case 2 when added more tables

The biggest advantage of using our MQO algorithm is that it can tackle the incoming queries at run time from the user. Consider a scenario where a user has submitted a batch of queris and the MQO generated a single shared plan for it. In the mean time, user submits more queries from the same pool of tables as the actual MQ have. We can fix new incoming quries into the already generated plan by passing the graph file and the pool of unsatisfied new queries to MQO. If the graph does not satisfy any of the queries, Case 1 will eventually run adding more JOIN operators to fit the queries in the graph. Now consider the second case if the new incoming queries consists of some new tables than the original MQ have but the boundry conditions are satisfied. In this case, Case 2 will run adding more tables on top of the graph and still able to generate a new plan at run time.

# Chapter 6

# Conclusion

In this paper, we have discussed about Multi Query Optimization which is an extended version of Shared Workload Optimization [1]. The main focus of the study was to generate a single plan with maximum sharing and least join cost. The join plans were implemented in GLADE [7]. In addition, we have provided an interface to submit SQL queries into GLADE, which are converted into graph and waypoint files executable on GLADE nodes. The greedy approach of cost based query optimization is used to generate optimal join plans which are then translated into GLADE execution plans, which runs on the GLAs. User can submit a pool of queries, which are then converted into batches or Multi Queries. Each Multi Query is parsed and optimized to produce a single JOIN plan, which is then compiled and executed in GLADE to return the results to the user for a batch of queries at the same time. The biggest achievement of the algorithm is its tendency to incorporate more queries from the user at run time while running the execution plan for the previously submitted batch of queries.

The testing was performed on the master cluster of GLADE and the benchmark used was TPCH. For our testing purposes, we have considered only simple SELECT-FROM-WHERE queries for generation of GLADE plans for both single and Multi Queries and also for the generation of graph and waypoint files. For future work, the same approach can be extended for the case of aggregate queries and queries involving GLAs.